\documentclass[]{interact}

\usepackage{epstopdf}% To incorporate .eps illustrations using PDFLaTeX, etc.
\usepackage{subfigure}% Support for small, `sub' figures and tables

\usepackage{natbib}% Citation support using natbib.sty
\bibpunct[, ]{(}{)}{;}{a}{}{,}% Citation support using natbib.sty
\renewcommand\bibfont{\fontsize{10}{12}\selectfont}% Bibliography support using natbib.sty
\usepackage[acronym]{glossaries} 
\usepackage{multirow}
\usepackage{multicol}
\usepackage{subfigure}	% Allow subfigures
\usepackage{adjustbox}
\usepackage{amssymb}
\usepackage{adjustbox}
\usepackage{eurosym}
\usepackage{algorithm}
\usepackage{algpseudocode}
\usepackage{pifont}
\newcommand{\xmark}{\ding{55}}%
\usepackage{color}
\usepackage{tikz}
\usepackage{array}
\usepackage[hidelinks]{hyperref}
\usepackage{tcolorbox}
\newcolumntype{M}[1]{>{\centering\arraybackslash}m{#1}}

\theoremstyle{plain}% Theorem-like structures provided by amsthm.sty
\newtheorem{theorem}{Theorem}[section]

\theoremstyle{definition}
\newtheorem{definition}[theorem]{Definition}

\theoremstyle{remark}

\newacronym[firstplural = Railway Undertakings (RUs)]{RU}{RU}{Railway Undertaking}
\newacronym{IM}{IM}{Infrastructure Manager}
\newacronym{EU}{EU}{European Union}
\newacronym{ACP}{ACP}{administrative and collaborative procedures}
\newacronym{CBA}{CBA}{social cost-benefit analysis}
\newacronym{WTP}{WTP}{Willingness-to-pay}
\newacronym{TSA-OPEN}{TSA-OPEN}{Time Slot Allocation in Open railway markets}
\newacronym{TAP}{TAP}{Track Allocation Problem}
\newacronym{TTP}{TTP}{Train Timetabling Problem}
\newacronym{ILP}{ILP}{Integer Linear Programming}
\newacronym{MILP}{MILP}{Mixed-Integer Linear Programming}
\newacronym[firstplural = Research Questions (RQs)]{RQ}{RQ}{Research Question}

\begin{document}

\articletype{}

\title{The time slot allocation problem in liberalised passenger railway markets: a multi-objective approach}

\author{
\name{Nikola Bešinović\textsuperscript{a}, Ricardo García-Ródenas\textsuperscript{b}, María Luz López-García\textsuperscript{b}, Julio Alberto López-Gómez\textsuperscript{c \dag}\thanks{\dag  Corresponding Author }, José Ángel Martín-Baos\textsuperscript{b}}
\affil{\textsuperscript{a}Technische Universität Dresden, Dresden, Germany; \textsuperscript{b}Escuela Superior de Inform\'atica, Universidad de Castilla-La Mancha, Ciudad Real, Spain; \textsuperscript{c}Escuela de Ingeniería Minera e Industrial de Almadén, Universidad de Castilla-La Mancha, Almadén, Spain}
}

\maketitle

\begin{abstract}
The liberalisation of the European passenger railway markets through the European Directive EU 91/440/EEC states a new scenario where different Railway Undertakings compete with each other in a bidding process for time slots. The infrastructure resources are provided by the Infrastructure Manager, who  analyses and assesses the bids received, allocating the resources to each Railway Undertaking. Time slot allocation is a fact that drastically influences the market equilibrium.  In this paper, we address the time slot allocation problem within the context of a liberalized passenger railway market as a multi-objective model. The Infrastructure Manager is tasked with selecting a point  from the Pareto front as the solution to the time slot allocation problem. We propose two criteria for making this selection: the first one allocates time slots to each company according to a set of priorities, while the second one introduces a criterion of fairness in the treatment of companies to incentive competition. The assessment of the impact of these rules on market equilibrium has been conducted on a liberalized high-speed corridor within the Spanish railway network.
\end{abstract}


\section{Introduction}
\label{sec:Intro}

In the last years, the European Directive EU 91/440/EEC has produced a significant change in the European passenger railway markets. This directive aims to facilitate the adoption of the Community railways to the Single Market and to increase their efficiency. To do that, it is intended to separate the infrastructure management from the provision of transport services and ensure access to the networks of the Member States for Railway Undertakings. The European so-called 4th railway package (EU Regulation 2016/2338) aims to remove the remaining barriers to the creation of a single European rail area. Additionally, the updated legislation includes the proposal to open up domestic passenger railways to new entrants and services from December 2019. This way, Railway Undertakings will be able either to offer competing services, such as a new train service on a particular route, or to bid for public service rail contracts through tendering.

The objectives of competition in the railway industry encompass several key aspects, including fare reduction (\cite{Vigren2017}), enhanced service frequencies (as discussed in \cite{Tomes2016}), and improved timetables (\cite{Dewei2019}, \cite{Yang2023}), among other factors. These efforts collectively aim to stimulate higher demand for railway services.

Two common market integrations are present: vertical and horizontal. The former represents a single operator (the so-called {\it incumbent operator} or {\it national railway undertaking}) that operates in a monopoly situation and assumes a captive demand, which was customary in the member states of the European Union. The latter, also known as open passenger railway market, new railway operators (the so-called entrants) compete with each other for the time slot to which rail services are allocated. However, the expected transition from a vertical to a horizontal market has been very difficult (\cite{Nash2008}, \cite{Sanchez2010}). Some of the barriers that entrants face and that jeopardise fair competition among operators are: inferior resources with respect to the incumbent operator or the lack of regulations to ensure fair competition, among others (\cite{Ristic2022}).

In a liberalised passenger railway system, there are three main interrelated actors: the \gls*{IM}, the \glspl*{RU} and the demand, which is represented by passengers. The \gls*{IM} is the owner and the manager of the railway infrastructure. The \glspl*{RU}, also known as train operating companies, are the companies that offer rail services to passengers using the infrastructure of the \gls*{IM}. Finally, the demand corresponds to passengers who want to make use of rail services to satisfy their transport needs for reaching from their origin to destination. The dynamics in a liberalised passenger railway system is the following: (i) the \gls*{IM} assigns capacities, which means a maximum number of time slot in which to operate, to each \gls*{RU}; (ii) the \glspl*{RU}, taking into account their assigned resources, place their bids to the \gls*{IM} in order to obtain a set of time slot in which to operate; (iii) the \gls*{IM} allocates time slot to the \glspl*{RU}, taking into account the bids received and the capacities assigned; and (iv) passengers choose one trip or another according to the final offer. 

%In order to ensure fair competition, there is an actor whose behaviour and decisions are crucial to achieve real competition.

The dynamics described above define a complex system of relationships between the actors, in which the particular interests of each of them conflict with the optimal performance of the system. The \gls*{IM} has a central role in the whole system, as their decisions to allocate time slots to different \glspl*{RU} may affect the equilibrium of the system and even lead to a monopoly situation. It is a critical process in order to guarantee fair competition between the \glspl*{RU}, since an unfair allocation or one that benefits the incumbent operator may affect the profitability and performance of the other \glspl*{RU} and the whole system (\cite{Broman2019}). 

In this context, there is a need to design optimal decision-making procedures in open passenger railway markets which guarantee equal treatment for all \glspl*{RU} involved in order to improve the socially efficient use of the railway system. It motivates the purpose of this paper: study how the decisions made by the \gls{IM}  when allocating time slots have an impact on the equilibrium state and determine the benefit of \glspl{RU}.

Time slot allocation can be addressed using different approaches (\cite{Bro2022}): to carry it out, the \gls*{IM} can use an auction-based criterion or set different prices for each time slot in order to encourage certain frequencies. In this work, the latter situation is analysed with the aim of considering how the \gls*{IM} should act when all companies are willing to pay such an amount and the role of \gls*{IM} is that of an arbiter of bids rather than a selfish role of maximising its own profit. 

This work proposes a new framework for \gls*{TSA-OPEN} problem in markets where multiple \glspl*{RU} operates under competition. For this purpose, \gls*{TSA-OPEN}  problem is considered within an equilibrium model. Thus, for each of the \glspl{RU} in the market, we obtain what their bid for operating rights in the time slots will be. This bid is so-called ``strategy'' and can be made up, in turn, of different requests or bids that the company can make with a given probability. The \gls*{TSA-OPEN} problem is formulated as a multi-objective optimisation problem as the \gls*{IM} has to deal with the conflicting interests of the different \glspl*{RU}. The canonical objective function of this problem is the profit function of the \glspl*{RU}. However, it remains unknown because its computation would require access to confidential data from them. In order to propose a criterion that can be shared by all participants, the distance between the time slot requested and the time slot allocated is transformed into a criterion that is transparent and known to all participants. The solution of this problem leads to multiple admissible solutions on what is known as the Pareto front.  However, although any of these solutions is acceptable, the market equilibrium situation varies considerably. Two different criteria are proposed for the selection of such solutions by the \gls*{IM}: (1) the \gls*{IM} allocates time slots to each \gls*{RU} according to a set of priorities, (2) the allocation is made on the basis of a fairness criterion in order to treat all \glspl*{RU} equally. We conducted computational experiments to assess the real impact of the decisions made by the \gls*{IM} on the system. The scenario under analysis is the liberalized high-speed corridor in the Spanish railway market, which connects Madrid and Barcelona. We examined the criteria mentioned above while assuming that all features of \glspl{RU} are equal, and our findings demonstrated significant differences based on the applied criterion.

%To this end, the computational experiments are based on a scenario in which all factors remain constant in order to study the real impact that the \gls*{IM}'s policies and decisions have on the system.

%\textbf{ADD: Main contributions of the paper (3-5 bullet points)}
The following are the main contributions of this article:
\begin{itemize}
    \item A mathematical multi-objective optimisation formulation of the time slot allocation problem for multiple \glspl*{RU} is provided. This formulation places the role of the \gls*{IM} at the centre as the arbiter in the deployment of the competing market.    
    \item Two selection criteria are introduced to assign the time slot: the use of priorities in the treatment of \glspl*{RU} and the application of an equity rule to treat all of them equally.
    \item An equilibrium model is proposed in order to integrate \gls{TSA-OPEN} into the bidding process of \glspl{RU} to achieve a market equilibrium. 
    \item \gls{TSA-OPEN} problem using the two proposed selection criteria are solved by using heuristic and exact approaches and their performance is compared. 
    \item The experiments conducted in this paper provide numerical evidence of how the criterion used by the \gls*{IM} significantly influences the equilibrium state, consequently affecting the economic performance of the \glspl*{RU}.   
    \item As a result of the above, the proposed framework constitutes a first fair approach to solve \gls*{TSA-OPEN}. The results underline the importance that the criteria employed by \gls*{IM} must be known by the \glspl*{RU}.

\end{itemize}

Finally, the rest of the paper is organised as follows. Section \ref{sec:Relwork} reviews the literature related to the time slot allocation problem. Then, section \ref{sec:Meth} describes the methodology employed in this work, including the necessary background about liberalised passenger railway systems and the formulation of \gls*{TSA-OPEN} problem. Later, section \ref{sec:Solapp} describes the  heuristics  to solve the \gls*{TSA-OPEN} problem. Section \ref{sec:NumExp} shows the experiments carried out and the results obtained in a high-speed corridor in the Spanish railway network. Finally, section \ref{sec:Conclusions} summarises the conclusions and further works derived from this work.

\section{Related work}
\label{sec:Relwork}
In this section, the most recent methods for time slot allocation in classical/closed/monopolistic and open/liberalised railway markets are discussed. Then, the reviewed methods for liberalised passenger railway markets are compared to the proposed time slot allocation methods of this paper. Summarily, Table~\ref{tab:state_art_comparison} shows this comparison. Finally, the scientific gaps are stated.

\subsection{Classical railway market}
Before the appearance of the EU regulations, which opened the door to the liberalisation of the European railways markets, the most common organisation of them was a completely integrated vertical and horizontal structure (\cite{Ali2020b}). It means there was a single actor which had a monopoly on the system, acting as \gls*{IM} and \gls*{RU} (\cite{Ali2019}). This way, the time slot allocation in this type of system was closed, opaque and carried out internally by the actor who monopolised the market (\cite{Ali2020b}).

The absence of competition in this market structure meant the aim of the time slot allocation problem was to build a feasible train timetable which maximised the profit of the company (\cite{Caprara2007}) and not violate track capacities and other operational constraints. In this kind of market, this problem is called the \gls*{TAP} or \gls*{TTP}. More formally, \cite{Schlechte2012} formulated this problem in the following way: given a macroscopic railway model and a set of train slot requests, the solution of \gls*{TTP} will be the subset of train requests which will be addressed as well as their exact arrival and departure times of the trains which will provide the service. 

The most traditional approaches to address this problem in single-track railway networks are \gls*{ILP} and \gls*{MILP} models (\cite{Carey1995}, \cite{Higgins1996}). In order to solve the problem, a wide variety of methods have been applied. Due to the fact that \gls*{TTP} is an NP-hard problem, solution approaches based on heuristics have been widely applied, such as greedy heuristics (\cite{Cai1998}), tabu search and genetic algorithms (\cite{Higgins1997}). 

\gls*{TTP} problem in general railway networks has also been traditionally addressed by means of graph theoretic formulations (\cite{Caprara2002},\cite{Caprara2006}). Departing from these works, many other modifications have been proposed, leading to \gls*{ILP} models and Lagrangian relaxed ones which are also solved by means of heuristic algorithms (\cite{Cachiani2008}). Interested readers can consult \cite{Lusby2011} for more information on such models and solutions.

\subsection{Liberalised railway market}

Under the new regulation, state-owned companies (incumbents) have given rise to the \gls{IM} and \glspl{RU}. In the EU, two models have emerged. The first involves splitting the incumbent company into two independent entities —one focused on network management and the other on transportation. This scenario is exemplified by the Spanish and Italian railway markets. The second structure entails replacing the incumbent company with a holding company comprising two separate entities. France and Germany are instances whose railway markets follow this arrangement. The market structure's impact is pivotal in fostering competition. Table~\ref{tab:open markets} lists the names of the companies in these railway markets.

\begin{table}[!h]
\centering
\resizebox{10cm}{!}{
\begin{tabular}{m{2cm}lll}
\toprule
\bf Country & \bf  Railway undertakings & \bf Infrastructure Manager\\ 
\toprule
Spain & RENFE, OUIGO, IRYO & ADIF\\
Italy & Trenitalia, Italo & RFI\\
France & SNFC VOYAGEURS, Trenitalia & SNFC Réseau\\
Germany & DB & DB Netz\\
\hline
\end{tabular}
}
\caption{Examples of open markets}
\label{tab:open markets}
\end{table}

In liberalised or open-access railway systems, competition marks the development of models to represent this type of market. Thus, the models built must take into account the conflicting interests between the \gls{IM}, the different \glspl*{RU} and also the interests of passengers in order to maximise the profit of all actors involved and contribute to social welfare. This way, the existing approaches are more focused on passenger's behaviour and on the fair allocation of time slots to the \glspl{RU} regarding the classic models, which are mainly focused on railway operations.  

\gls*{TSA-OPEN} is a key problem to ensuring fair competition in competitive and open-access railway markets. It can be broadly defined as follows: given a set of \glspl*{RU}, a set of time slots, the capacities of each \glspl*{RU} to operate them and the bids of the different \glspl*{RU} to operate the time slot, the result of \gls*{TSA-OPEN} will be an allocation of the time slot to the different \glspl*{RU}. According to \cite{Bro2022}, fairness is one of the most important features that a \gls{TSA-OPEN} method must meet. It is the task of \gls*{IM} to allocate time slots to different \glspl*{RU} taking into account the capacities of each one and solving the conflicts between them. A conflict or conflict state occurs when two or more \glspl*{RU} apply for access to the same track at the same time (\cite{Schlechte2012}). In open-access competitive railway markets, conflict situations are frequent, since they are encouraged by the number of \glspl*{RU} in the market and the amount of traffic on the network among others (\cite{Bro2022}). 

Traditionally, the time slot allocation problem has been solved through negotiations between the \gls*{IM} and the \glspl*{RU}, voluntary agreements and other simple criteria (\cite{Stoja2019}). However, these methods are not very effective in competitive and open-access markets (\cite{Bro2022}). 
Currently, the main methods to solve the time slot allocation problem are grouped into three main approaches (\cite{Bro2022}): based on \gls*{ACP}, approaches based on \gls*{CBA} and allocation based on \gls*{WTP}. 

The first approach for time slot allocation methods is based on \gls{ACP}. This approach is used in all \gls*{EU} countries and consists of applying a set of simple and pre-defined criteria to solve conflicts. Concerning this approach, \cite{Bro2022} describes the case of the Belgium railway market, where the \gls*{IM} classifies the train services according to priorities and when the conflict occurs, the least-priority class of service is adapted. Another recent example is \cite{Trafi2020}. In this document, the time slot allocation procedure of the Swedish railway market is described. In it, the different \glspl*{RU} submit their time slot request to the \gls*{IM} according to the national network statement. Then, the \gls*{IM} makes the allocation solving minor conflicts by means of small adjustments. Major conflicts are solved by means of informal discussions between the involved \glspl*{RU}. If an agreement is not reached, the \gls*{IM} solves the conflict using a set of rules of pre-defined criteria.  The main advantages of administrative and collaborative procedures are usually that they are very clear and simple (\cite{Bro2022}). However, their main drawbacks are related to their problems of treating all companies fairly. 

\gls*{CBA} is the second main approach for solving the time slot allocation problem. It is a very well-known methodology in the transportation field. However, the literature applying this approach to the time slot allocation problem is scarce. The fundamentals of \gls*{CBA} consist of defining all costs and benefits of the time slot allocation procedure in economical terms and summing up in order to know the extent to which the method designed contributes to welfare. Because of this, the main characteristic of a \gls*{CBA} time slot allocation procedure is fairness. However, the need to recover enough data from the system (\glspl*{RU} and passengers) to design it, could affect the achievement of the remaining good properties. Recently, \cite{Ali2020} proposed a \gls*{CBA} based method for allocation and timetable comparison. For further reading, readers are referred to \cite{Bro2022}, where their authors propose a mixed method which combines administrative procedures and \gls*{CBA} allocation. 

Finally, \gls*{WTP} methods constitute the last approach to solving the time slot allocation problem. It is a very common approach, especially in competitive and open access markets, where multiple \glspl*{RU} compete between them in an auction or bidding process. This way, the \glspl*{RU} are sorted according to how much money they are willing to pay for a time slot. The main advantages of \gls*{WTP} methods are usually their simplicity and manageability. However, these methods normally suffer from a lack of transparency and above all from a lack of fairness. There are many efforts in literature to develop fairer \gls*{WTP} methods based on auctions. Some of these works are \cite{Kuo2015}, where a combinatorial auction was employed in ongoing allocation schemes,  \cite{Talebian2018}, which proposes a negotiation approach in vertically integrated railway systems employing game-theoretic concepts and \cite{Stojad2019}, where an algorithm for iterative time slot allocation is proposed under a scheme of hybrid auctions. Finally, another related example is \cite{Stoja2019}, where their authors propose an allocation algorithm based on auctions to make the allocation. 

\begin{table}[!h]
\centering
\resizebox{15cm}{!}{
\begin{tabular}{m{2cm}llm{3cm}lcm{2.5cm}m{2.5cm}cc}
\toprule
\multicolumn{1}{c}{Reference} &  & \multicolumn{2}{c}{Railway System}        &  & \multicolumn{5}{c}{Framework}  \\
\cmidrule(lr){1-1} \cmidrule(lr){3-4} \cmidrule(lr){5-10}
\multicolumn{1}{c}{}                           &  & Country     & Type of transport                    &  & Approach                    & Model                 & Method   & Implementation               & Fairness    \\
\cmidrule(lr){1-1} \cmidrule(lr){3-4} \cmidrule(lr){5-10}
\cite{Bro2022}            &  & Belgium     & Freight and passengers              &  & \gls*{ACP}    & Rule-based    & Priorities  & Theoretical & \xmark        \\
\hline
\cite{Trafi2020}             &  & Sweeden     & Freight and passengers              &  & \gls*{ACP}    & Rules and negotiation-based    &  Agreement and priorities & Theoretical       & \xmark      \\
\hline
\cite{Ali2020}               &  & Sweeden     & Passengers (commuter-trains) &  & \gls*{CBA}        & Mathematical modelling & Computing loss of societal benefits & Simulation & \checkmark  \\
\hline
\cite{Stoja2019}            &  & European    & Passengers              &  & \gls*{WTP} & Negotiation     & Agreement and auction  & Theoretical & \xmark    \\
\hline
\cite{Kuo2015}              &  & Synthetic   & Freight &  & \gls*{WTP} & Auctions Modelling   & Optimisation & Simulation   & \xmark        \\
\hline
\cite{Talebian2018}           &  & Synthetic   & Freight and Passengers &  & \gls*{WTP}        & Game-Theory: Equilibrium problem & Optimisation & Simulation & \checkmark \\
\hline
\cite{Stojad2019}             &  & European    & Passengers &  & \gls*{WTP} & Mathematical modelling & Hybrid auctions & Simulation      & \xmark    \\
\hline
\cite{Bro2022}            &  & Synthetic & Passengers                    &  & Hybrid & \gls{ACP} and \gls{CBA} & Welfare losses & Theoretical & \checkmark \\
\hline
This paper: priority and equity rules &  & Spanish    & Passengers &  & \gls{ACP} & Mathematical modelling & Optimisation & Simulation  & \checkmark \\
\hline
\end{tabular}
}
\caption{Comparison of state-of-the-art works related to time slot allocation.}
\label{tab:state_art_comparison}
\end{table}

Table~\ref{tab:state_art_comparison} shows a comparison of the discussed works related to time slot allocation in liberalised passenger railway markets. Derived from this table, it is possible to conclude the following: first, there are many works which propose theoretically time slot allocation procedures, there is therefore a need to implement and validate the proposed allocation methods experimentally. Second, the references indicated in the table as fair do not always guarantee equal treatment among all companies. This is because, when the allocation is based on \gls*{WTP}, the allocation may be unfair as not all \glspl*{RU} have the same resources. On the other hand, when the allocation is based on \gls*{CBA}, the study of these indicators is not always possible as it requires data that is often not available and it may favour the incumbent operator and make it more difficult for new entrants to compete. This way, there is a gap in designing allocation approaches that ensure equal treatment of all companies and that are simple and easy to implement. 
%(Explicitly define existing gaps in the current literature: against the current closed market and open market approaches; also related to methodology used for solving the problem.)
%First, ...
%Second, ...

In this paper, we focus on the liberalized Spanish railway market as our reference case. This market presents challenges for the application of \gls*{WTP}-based approaches due to the absence of auction mechanisms employed by the \gls{IM}. Additionally, the \gls*{CBA} approach faces difficulties due to the requirement of reliable socio-economic data, limiting its transparency. Instead, we adopt an \gls*{ACP}-based approach. 

Notably, an aspect that remains unexplored in the existing literature is the assessment of the impact of \gls{TSA-OPEN} on competitive dynamics. To address this gap, we introduce two criteria that offer distinct treatments of the \glspl{RU}. The priority criterion reflects the strategic dominance situation of the incumbent, while the fairness criterion advocates for equal treatment of all \glspl{RU}. Both of these developed methods are characterized by their simplicity and practicality in terms of design and implementation. This study aims to shed light on the implications of these criteria for competition in the railway market.

\section{The time slot allocation  problem in open railway markets }
\label{sec:Meth}

In this section, a mathematical formalization of \gls*{TSA-OPEN} is provided. To achieve this, the foundation of a liberalized railway market is initially presented. Subsequently, the \gls*{TSA-OPEN} problem is formulated as a multi-objective optimization problem. The objective of the \gls{IM} is to select one of the solutions from the Pareto front to be implemented in the system. In this article, two criteria are proposed for choosing a solution from the Pareto front: a priority-based rule and an equity rule. The first criterion can be viewed as administrative and collaborative procedures based on priorities, while the second is designed to promote equity in the treatment of the \gls{RU}.

\subsection{Liberalised Railway Markets}
\begin{figure}[!h]
	\centering
	\includegraphics[scale=0.30]{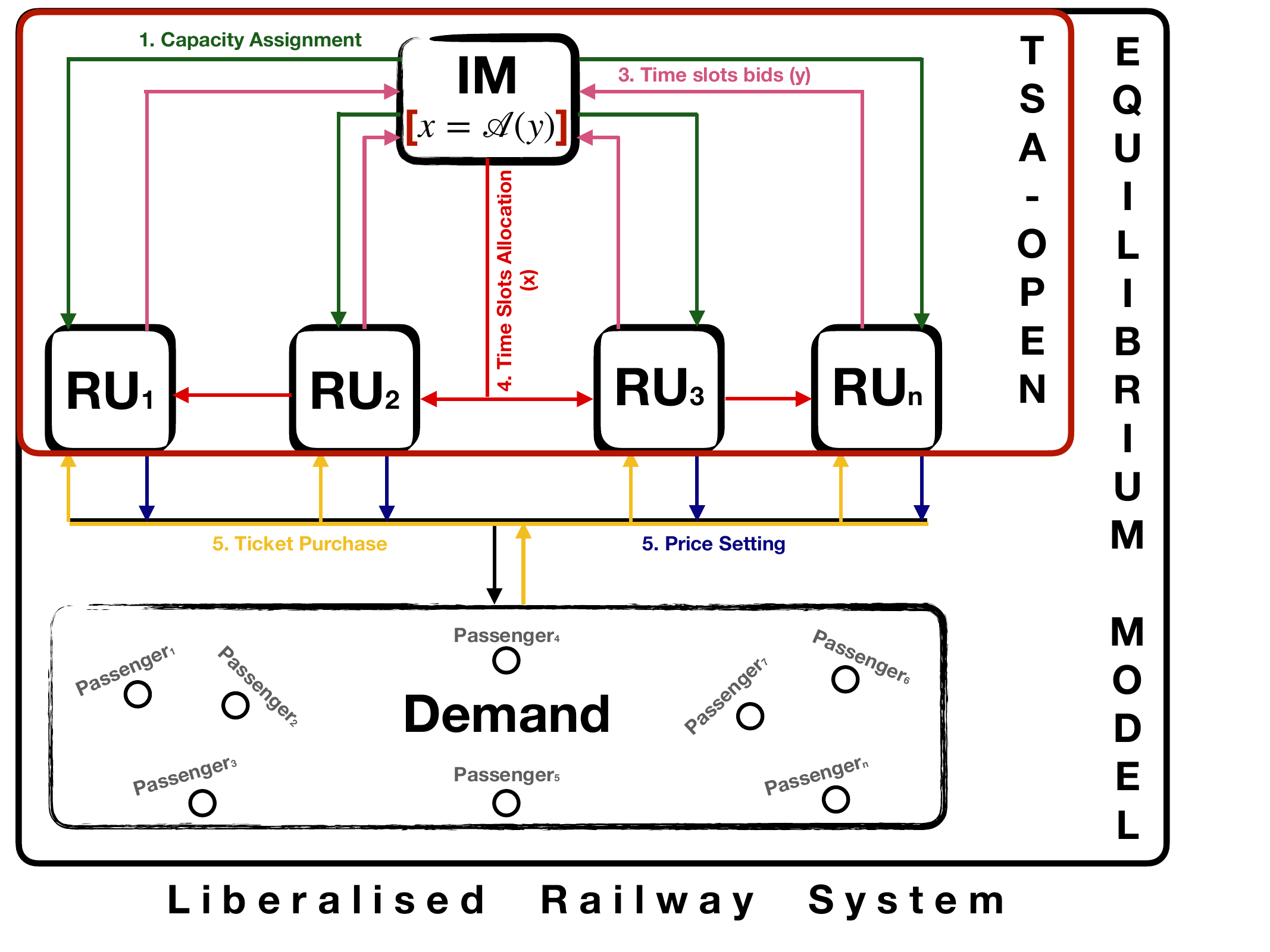}
    \caption{General scheme of a liberalised railway system and time slot allocation problem.}
	\label{fig:Railway_Sistem}
\end{figure}

Figure~\ref{fig:Railway_Sistem} shows a scheme of a  liberalised passenger railway market stated on the basis of the Spanish case. The railway system is composed of three main actors: a public company, the so-called \gls*{IM} that provides the infrastructure  and a set of private companies. The second one is \glspl*{RU}, that operate on the railway network. A specific \gls*{RU} is denoted by $o$ and the set of \glspl*{RU} by means of the set $\mathcal{O}$. Finally, the third main actor is the demand, which is represented by passengers. In a liberalised or open railway market, an equilibrium situation, in terms of the profit between \glspl*{RU}, is desirable in order to promote competition, enhance the service frequencies and reduce fares. 

The \gls*{IM} selects one or several corridors for liberalization. The primary responsibility of the \gls{IM} is to create a set of packages, each characterized by its capacity, which will be auctioned to a group of \glspl{RU}. Based on the various offers received and technical considerations, the \gls{IM} allocates these packages among a subset of \glspl{RU} that have submitted bids. The capacity of the package assigned to \gls{RU} $o$ is represented as $k_o$. This corresponds to the fraction of the time slots that each \gls{RU} can apply for in the auction process. Therefore, this company may operate in up to $k_o|{\cal R}|$ time slots, where $|{\cal R}|$ represents the total number of time slots in the corridor. These capacity assignments restrict the requests of each \gls{RU}. Capacity allocation is the initial factor that influences whether fair competition exists among the \glspl{RU}.

A time slot, represented as $r$, is linked to a train route, including its arrival and departure times at stations. To be more precise, it is a segment on a space-time diagram where a train operates without conflicts.

Once the capacity assignment is carried out, the \glspl*{RU} are ready to make their bids to the \gls*{IM} in order to obtain the operating rights in a time slot. To do that, the \glspl*{RU} need to take into account the capacities that have been previously assigned to them. The bids are modelled by a vector of binary variables, whose length is equal to the number of time slots. This way, a bid made by $o$ company is denoted by $y_o$ vector. The value of each component of the vector $y_o$ is equal to $1$ when the corresponding time slot is requested by \gls*{RU} $o$ and $0$ otherwise. Equation~(\ref{eq:bids_yo}) shows this formalisation. 
\begin{equation}
\label{eq:bids_yo}
    y_o = [y_{or_1},y_{or_2},...,y_{o|\mathcal{R}|}].
\end{equation}

Note that for the binary variable $y_o$ to be feasible, it must satisfy the capacity restriction modelled by Equation~(\ref{eq:capacity_constraint}) which prevents the \gls{RU} $o$ from bidding for more time slots than it has capacity allocated to it.
\begin{equation}\label{eq:capacity_constraint}
    \sum_{r \in {\cal R}} y_{or} \le k_o |{\cal R}|.
\end{equation}

After the bids placement, the \gls*{IM} allocates the time slots to each \gls{RU}. To do that, the \gls{IM} resolves conflicts present in the request $y_o$ concerning other \glspl{RU}, transforming it into a conflict-free request denoted as $x_o$. A conflict occurs when a time slot is requested by more than one \gls{RU}. The result of the allocation process for \gls*{RU} $o$ is the allocation vector $x_o$ shown in Equation~(\ref{eq:alloc_xo}). It must be noted that the value $1$ in an element of $x_o$ means that the corresponding time slot has been assigned by the \gls*{IM} to the company $o$. Otherwise, the value in that element will be $0$. 
\begin{equation}
\label{eq:alloc_xo}
    x_o = [x_{or_1},x_{or_2},...,x_{o|\mathcal{R}|}].
\end{equation}

In summary, the \gls{IM} receives a feasible bid $y$ of \glspl{RU}, and then \gls{IM} will provide an allocation of time slot $x$ free of conflicts. Equation~(\ref{eq:compact_tsa_open}) abstracts this procedure as a mapping ${\cal A}$.
\begin{equation}\label{eq:compact_tsa_open}
    x= \mathcal{A}(y).
\end{equation}

\subsection{\gls*{TSA-OPEN} modelling}

One of the most important objectives of \gls{IM} is to provide a fair allocation between the \glspl{RU} involved in the market. Fairness must be a key criterion if it is intended to promote the development of competition between \glspl{RU} in the market, without excluding any of them, and thus offering passengers better travel alternatives. 

The \gls{IM} acts as an arbitrator, who allocates the time slot considering the specific objectives of each \gls{RU}. The profit or payoff of \gls{RU} $o$ is represented by the function $F_o({x}_o,{x}_{-o})$, where $x_o$ represents the allocation to \gls{RU} $o$ and $x_{-o}$ represents the allocation to the rest of the \glspl{RU}. Therefore, the \gls{IM} should keep in mind the following multi-objective problem in the definition of allocation method $\cal A$.

\begin{align}
     \hbox{maximize} \; &F=[F_1({x}_1,{x}_{-1}),\ldots,F_o({x}_o,{x}_{-o}),\ldots, F_{|{\cal O}|}({x}_{|{\cal O}|},{x}_{-{|{\cal O}|}})] \label{eq:multi_objective_problem}\\
     \hbox{subject to:}& \sum\limits_{r' \in {\cal R}}h^r_{or'} =   y_{or} \label{eq:const1} \\ 
     &\sum\limits_{r' \in {\cal R}}h^{r'}_{or} = x_{or} \label{eq:const2}\\ 
     & h^r_{or'} \in  \{0,1\}
    \label{eq:const3}
\end{align}  

Constraints (\ref{eq:const1})-(\ref{eq:const3}) represent the possibility of re-timing a time slot, introducing auxiliary variables $h^r_{or'}$. These variables take the value $1$ if the time slot $r$ requested by \gls*{RU} $o$ is re-timed to the time slot $r'$. 

If $\mathcal{P}$ represents the Pareto front of the problem (\ref{eq:multi_objective_problem})-(\ref{eq:const3}), the \gls{IM} should employ an allocation criterion that satisfies the condition:
$$ x= {\cal A} (y) \in {\cal P}.$$
Otherwise, there would exist an allocation decision that would be more beneficial for all \glspl{RU}. The \gls{IM} is unaware of the multi-objective problem due to the unavailability of economic data from the \glspl{RU}. To address this, a simplified and transparent version that can be understood by all involved parties is proposed. It is based on minimizing the discrepancies between the time slots requested by the \glspl{RU} and the allocations ultimately made by the \gls{IM}.

In this new formulation, the aim of each \gls*{RU} $o$ is to minimise its total deviation $D_o$, that is, the sum of the deviations of all the requested time slots. Formally, it is noted by Equation~(\ref{eq:deviation}).
\begin{equation}
    \label{eq:deviation}
    D_o=\sum_ {r\in {\cal R}} \delta_{or},
\end{equation}
\noindent where $\delta_{or}$ is the deviation of the rescheduled time slot $r$ requested by operator $o$ and is computed by the Equation~(\ref{eq:deviation_calculus}).
\begin{equation}
    \label{eq:deviation_calculus}
    \delta_{or}= \frac{1}{2}\sum_{r' \in {\cal R}} |  y_{or}-h^r_{or'}| |t_r-t_{r'}|,
\end{equation}
being $t_r$ the time instant of time slot $r$ and $t_{r'}$ is the time instant of the re-timed time slot $r'$.\\

The deviation-based formulation leads to the following multi-objective optimisation problem which will be addressed by the \gls{IM} to solve \gls{TSA-OPEN} problem: 
\begin{eqnarray}
\label{eq:multi_objective_problem_deviation}
     \hbox{maximize}  & D=[D_1,...,D_o,...,D_{|\mathcal{O}|}]\\
     \nonumber \hbox{subject to:} & (\ref{eq:const1})-(\ref{eq:const3}) 
\end{eqnarray}  

Solving \gls{TSA-OPEN} problem allows to obtain a solution from the Pareto front. The following subsections describe two different criteria or approaches to choose Pareto optimum solutions: the first is based on a priority rule approach and the second is based on an equity rule.

\subsection{TSA-OPEN modelling: priority rule}

The first criterion proposed in this paper to choose a Pareto optimum is the priority rule. In this approach, the requests of the \glspl{RU} are served according to a set of priorities that determine which \gls{RU} is served first, which second and so on. Usually, priorities are related to the capacity allocated to the \glspl{RU}. This way, the \gls{RU} with the highest capacity will be served first while the \gls{RU} with the lowest assigned capacity will be served last. 

The priority rule means that an iterative process is carried out. In each iteration, an allocation problem is solved for a \gls{RU} according to their priorities. This way, at the start of the resolution process, no time slot has been allocated to any \gls{RU}. Then, according to the priorities of each of the \glspl{RU}, the one with the highest priority will be selected and its requested time slot will be allocated, as none have been allocated previously. Later, the next \gls{RU} in order of priority will be selected and the allocation procedure will be repeated. However, in this case, it is possible that conflicts may occur and that the second \gls{RU} has requested time slots that have already been allocated to the first \gls{RU}, so in this case the \gls{IM} shall re-plan those slots trying to minimise the deviation from the requested time slot. This iterative process will be carried out for all the \glspl{RU}. As the allocation is carried out for a new \glspl{RU}, an increasing number of conflicts will arise. This is due to the additional slots allocated to previously served \glspl{RU}, being necessary to re-plan the time slot allocated to each one.

Let's assume that the \glspl{RU} have been prioritized in the order $o_1, \ldots, o_{|\cal O|}$, and suppose that all companies $\{ o1,\ldots.o'-1\}$ have been analyzed, calculating their respective allocation variables $x_{or}$. Then, for the time slot allocation for company $o'$, the \gls{IM} will minimize its deviation:
\begin{align}
\hbox{Minimize \,\,} \;& D_{o'} \label{eq:problem_priority} \\
\hbox{Subject  to:} & \sum_{r' \in {\cal R}}h^r_{o'r'} =   y_{o'r}, \label{eq:cons_replan_prior1} \ \ \ \forall  r \in {\cal R} \\
&\sum_{r' \in {\cal R}}h^{r'}_{o'r} = x_{o'r}, \ \ \ \forall r \in {\cal R}   \label{eq:cons_replan_prior2}\\
& \delta_{o'r}= \frac{1}{2}\sum_{r' \in {\cal R}} |  y_{o'r}-h^r_{o'r'}| |t_r-t_{r'}|, \label{eq:delta_prior} \ \ \ \forall r \in {\cal R} \\
& D_{o'}=\sum_ {r\in {\cal R}} \delta_{o'r} \label{eq:Dprior} \\
&  x_{o'r}+ \sum_{o \in \{o1,\cdots, o'-1\}} x_{or} \le 1, \ \ \ \forall r \in {\cal R} \label{eq:allocatedprior}  \\
&h^r_{o'r'}\in\{0,1\}, \label{eq:binaryreplan_prior} \ \ \ \forall r,r' \in {\cal R}\\ \nonumber
\end{align}
\noindent where the constraints (\ref{eq:cons_replan_prior1})-(\ref{eq:cons_replan_prior2}) models replanning time slots, the constraints (\ref{eq:delta_prior})-(\ref{eq:Dprior}) define the computation of the deviation for the \gls{RU} $o'$, and the constraint (\ref{eq:allocatedprior}) prevents a time slot from being allocated to more than one \gls{RU}. Note that the variables $x_{or}$ in Equation~(\ref{eq:allocatedprior}) have been computed in prior iterations and, in this formulation, they act as parameters. Finally, constraint (\ref{eq:binaryreplan_prior}) shows the binary nature of replanning variables. 

Priority rule modelling is an \gls{ACP} time slot allocation approach which, although simple and practical, represents the strategic dominance of the incumbent \gls{RU} in the market. In the next section, an equity rule approach is proposed in order to ensure a fair allocation to all the \glspl{RU} in the system.

\subsection{TSA-OPEN modelling: equity rule}

In contrast to the method introduced earlier, where the \glspl{RU} are treated unequally, we introduce a method to ensure a fairer allocation. The proposed criterion, named equity rule, tries to balance the deviations suffered between \glspl{RU}. In this case, the optimization model aims to minimise the total sum of deviations across all \glspl{RU} within the market. However, it also enforces a constraint to limit the maximum difference that can be observed between the deviations experienced by the \glspl{RU}.

The equity criterion is related to the capacity assigned to each \gls{RU}. In this paper, the proposed approach imposes that the total displacement for the time slot of the \gls{RU} $o$ must be proportional to the size of its operating capacity, which is noted by $D_o = k_o\Delta$. However, due to the discrete nature of this constraint, it is not always possible to achieve exact equality in maintaining this proportionality. To address this limitation, a tolerance parameter $\varepsilon$ is introduced to ensure its satisfaction.

Given the time slot requests $y$ by the \glspl{RU}, we calculate the total number of requested time slot as $n_y = \sum_{o \in \mathcal{O}}\sum_{r \in \mathcal{R}}y_{or}$, and the \gls{IM} formulates the TSA-OPEN problem under the equity rule as (\ref{eq:problem_equity}):
\begin{align}
\hbox{Minimize \,\,} \;& \sum_{o \in \cal  O} D_o\label{eq:problem_equity}\\
\hbox{Subject  to:} & \sum_{r' \in {\cal R}}h^r_{or'} =   y_{or}, \hspace{0.5cm}\forall  o \in {\cal O},  \forall r \in {\cal R}  \label{eq:cons_replan_eq1}\\
&\sum_{r' \in {\cal R}}h^{r'}_{or} = x_{or},\hspace{0.5cm}\forall  o \in {\cal O},  \forall r \in {\cal R}  \label{eq:cons_replan_eq2} \\
& \delta_{or}= \frac{1}{2}\sum_{r' \in {\cal R}} |  y_{or}-h^r_{or'}| |t_r-t_{r'}|, \hspace{0.5cm}\forall  o \in {\cal O},  \forall r \in {\cal R} \label{eq:delta_equi}\\
& D_o=\sum_ {r\in {\cal R}} \delta_{or}, \hspace{0.5cm}\forall  o \in {\cal O}\label{eq:Dequi}\\
& \sum_{o \in {\cal O}} x_{or} \le 1, \hspace{0.5cm} \forall r \in {\cal R}\label{eq:allocatedequi}\\
&\Delta = \frac{1}{n_y}\sum_{o \in {\cal O}}  \sum_{r\in{\cal R}} \delta_{or} \label{eq:delta_def}\\
& \Delta  -\varepsilon \le \frac{1}{k_o\cdot n_y} D_o \le  \Delta  +\varepsilon, \hspace{0.5cm}\forall  o \in {\cal O}\label{eq:integer_constraint}\\
&h^r_{or'}\in\{0,1\} \hspace{0.5cm}  \forall r,r' \in {\cal R},\forall  o \in {\cal O} \label{eq:binary_replan_equi}\\ \nonumber
\end{align}
\noindent where the constraints (\ref{eq:cons_replan_eq1})-(\ref{eq:Dequi}) and (\ref{eq:binary_replan_equi}) are similarly formulated as for the priority rule, but they apply to all \glspl{RU} collectively. Constraint (\ref{eq:allocatedequi}) ensure a time slot is only allocated to one \gls{RU}. The constraint (\ref{eq:delta_def}) computes the average deviation for the time slot and it is denoted by $\Delta$ and  the  constraint (\ref{eq:integer_constraint}) imposes that the average deviation for all \glspl{RU} is equal to $\Delta$ plus or minus an error  $\varepsilon$.

\section{Assessing \gls{TSA-OPEN} in liberalized markets: an equilibrium game}
\label{sec:Solapp}

The purpose of this article is twofold. First, it seeks to establish precise guidelines for ${\cal A}$ to ensure transparent and equitable allocation of time slots. Since the conceptual method ${\cal A}$ for assigning time slots affects the equilibrium of the open railway system, the second objective is to evaluate the impact of these guidelines on competition. To accomplish this, tools must be developed for assessing the rules $\cal A$. In this section, we introduce an equilibrium game approach as a model for the open railway market, in which the \gls{TSA-OPEN} $x = \mathcal{A}(y)$ is a component of this equilibrium model. 

The equilibrium problem at hand can be stated as a game in which the players are \glspl*{RU} and the strategies are the time slot requests $y$. The \glspl*{RU} need to determine which strategies  to employ and with what probability to request them from the \gls*{IM} in order to maximise their payoff functions. For example, let's suppose a \gls{RU} which has two request options: $y_1$ and $y_2$. This \gls{RU} is interested in knowing the probability values $p_1$ and $p_2$, such as $p_1 + p_2 = 1$, where $p_1 \geq 0$ and $p_2 \geq 0$, for playing these requests to maximize its profit. 

The payoff function $F$ corresponds to the economic revenue of the \gls{RU} which is computed as the difference between the revenues from ticket sales and the costs of operating the time slot together with the costs of acquisition and depreciation of rolling stock. 

The revenue obtained from ticket sales for the time slot $r$ can be computed by
$
    b_r = \sum_{\omega \in {\cal W}}   g_{r\omega} z_{r \omega},
$
where $g_{r\omega}$ is the demand of the time slot $r$ in the origin-destination pair $\omega$ and $z_{r\omega}$ is the ticket price of  for the  origin-destination trip  $\omega$ at time slot $r$. 
%The  Equation~(\ref{eq:J}) shows the computation of the revenue from ticket sales for \gls{RU} $o$.
The revenue from ticket sales for \gls{RU} $o$ is computed as
\begin{equation}\label{eq:J}
    J_o(x_o,x_{-o})= \sum_{r\in {\cal R}}  x_{or} b_r.
\end{equation}
Note that the expression (\ref{eq:J}) is a simplification of the problem, in which it is assumed the number of passengers attended by a service depends only on the allocated time slot.

Concerning cost for \gls*{RU} $o$, it consists of two components. The first part is associated with the operative costs and the second with investment costs to provide the established services. The operational costs $f_{or}$ take into account the network access variable  cost fixed by \gls*{IM} and the staff and operational costs by ticketing, advertisement, passenger attendance, drivers, and so on. 
Regarding investment costs, \gls*{RU} $o$ has to assume the cost of leasing rolling material to offer the service ($C_o$). Thus, the investment costs will depend on the number of units and the typology necessary to deploy the service. It is assumed all trains have the same typology and the rolling stock costs depend on the minimum number of trains needed to operate all accepted time slots.  In investment costs, the maintenance costs are also included as well as the depreciation costs of the rolling stock. 
It is considered that  the fixed cost to network access $c_a$ has already been incurred by \glspl*{RU}. It does not affect the choice of time slot but is included to adjust the  balance sheet of \gls*{RU}. The minimum number of trains needed to operate the service network (time slot) is denoted  by $n_T(x_o)$. This introduces the investment needs of the operator and investment costs. 
Finally, the payoff function for a \gls{RU} $o$, given the time slot allocated to it ($x_o$) and the time slot allocated to the rest of \glspl{RU} ($x_{-o}$) is defined as% it is shown in Equation~(\ref{eq:F}):
\begin{equation}\label{eq:F}
    F_o(x_o,x_{-o}) = J_o(x_o,x_{-o})- \sum_{r\in {\cal R}_o}f_{or} x_{or}- C_{o} n_T(x_o) -c_a.
\end{equation}

Let's consider $\mathbf{p}_o=(\cdots, p_{y_o},\cdots )$ as a probability distribution over the set of strategies that the \gls{RU} $o$ can employ ($y_o \in { Y}_o$). Then, for a game strategy  $y=(y_{o_1} \cdots,y_{o_{|{\cal O}|}}) \in { Y}={ Y}_1 \times \cdots \times { Y}_{|{ O}|}$, its probability of being applied is given by:
$
    p_y=\prod_{s=1} ^{|{\cal O}|} p_{y_{o_s}},
$
and therefore, the expected benefit for \gls{RU} $o$ is:
\begin{eqnarray}
\label{eq:equilibrio1}
    u_o(\mathbf{p}_o,\mathbf{p}_{-o}) &=& \sum_{y \in { Y}} p_y F_o(x_o,x_{-o}),\\
    \label{eq:equilibrio2}
    x&=&{\cal A}(y).
\end{eqnarray}

In an equilibrium scenario, no \gls{RU} can unilaterally enhance their expected economic outcomes. We will now mathematically formulate this situation:
\begin{definition}[Nash equilibrium bid] A strategy profile $\mathbf{p}^*_{ {\mathbf{y}}}$ for the bid  is a Nash equilibrium if  for all $o\in {\cal O}$,  for all $\mathbf{p}_{ {\mathbf{y}}_o} $ and for all $y_o\in Y_o$
\begin{equation}
\label{eq:model}
    u_o(\mathbf{p}_{ {\mathbf{y}}_o},\mathbf{p}^*_{ {\mathbf{y}}_{-o}}) \le u_o \left (\mathbf{p}^*_{ {\mathbf{y}}} \right ).
\end{equation}
\end{definition}

It is worth noting that the allocation rule $\cal A$ impacts the profit $u_o(\mathbf{p})= \sum_{y \in Y} p_y F_o \left ( \,{\mathcal A}(y)\, \right)$ and, consequently, alters the equilibrium situation. This is the effect we aim to analyze.

Furthermore, it should be noted that calculating the equilibrium situation requires determining $x={\cal A}(y) \hbox{ for all } y \in Y$. The proposed models are based on integer linear programming\footnote{Note that the absolute value of a magnitude appearing in constraints (\ref{eq:delta_prior}) and (\ref{eq:delta_equi}) can be modelled using linear programming through the following trick:
\begin{eqnarray*}
|x|= x^+ + x^-\\
x=x^+ - x^-\\
x^+\ge 0, x^-\ge 0
\end{eqnarray*}
}. While commercial solvers efficiently handle these models, the set $Y$ presents a vast number of alternatives, necessitating the resolution of a substantial number of these problems. Consequently, to streamline the incorporation of $\cal A$ into an equilibrium computation method, it is advisable to employ heuristic algorithms for its calculation. In the following two subsections, we introduce such algorithms for each of the rules.

\subsection{Priority rule}\label{subsec:Priority}

In order to solve \gls*{TSA-OPEN} problem considering the priority rule modelled in section \ref{sec:Meth}, a heuristic algorithm is proposed.  In this method, the set of priorities that define the order in which the \glspl{RU} will be attended is considered to be given. For example, priorities may be directly related to the capacities of the \glspl*{RU} in such a way that the more capacity a \gls*{RU} has, the more priority it has. Thus, the \gls*{IM} starts to allocate time slots for the company which has the highest priority and so on. The allocation is done assuming that the previous time slot has already been allocated and re-planning the conflicting time slot to the nearest available one. To illustrate this solution procedure, Algorithm \ref{code:priority} shows the pseudo-code of a heuristic algorithm for \gls*{TSA-OPEN} problem based on the priority rule.   

\begin{algorithm}
\begin{algorithmic}[1]
\Require $\mathcal{O}, \mathcal{R}, y_1,y_2,...,y_{|\mathcal{O}|}$
\Ensure  $x$
\Function{\gls*{TSA-OPEN} using priority rule}{}
\State -- Priorities vector with the ordered indexes of the \glspl*{RU} to be served -- 
\State Priorities $= [o_1,o_2,o_3,...,o_|\mathcal{O}|]$
\State -- Solve an allocation problem for each \gls*{RU} --
\For{$i=1:|\mathcal{O}|$} 
\State -- Select \gls*{RU} by priority -- 
\State $o' = \mathcal{O}[Priorities[i]]$ 
\For{$r$ requested time slot by o} 
\If {$r$ is available}
\State -- Direct allocation of free time slot --
\State $x_{o'r} = y_{o'r} $ 
\Else
\State -- Re-timing of conflicting time slot  -- 
\State $h^{r'}_{o'r} =1$ where $r'$ is the closet available time slot of the requested $r$.
\State -- Allocate the re-timing time slot -- 
\State $x_{o'r'} =1$ 
\EndIf
\EndFor
\EndFor
\EndFunction
\end{algorithmic}
\caption{Pseudocode for solving heuristically \gls*{TSA-OPEN} problem using a priority rule}\label{code:priority}
\end{algorithm}

This algorithms departs from a vector of priorities defined in line $3$. It contains a priority value for each \gls{RU} which will determine the order in which the requests of each one will be processed with respect to the other \glspl{RU}. Then, in line $5$, a for loop is used to select, in the first place, the \gls{RU} with the highest priority and whose requests will be handled. Once the \gls{RU} to be served is determined, a nested for loop in line $8$ goes through the requests made by the \gls{RU}, so that if the requested time slot is free, it is directly allocated to that \gls{RU} (line $11$) and if it has already been allocated, the nearest available time slot is assigned (line $16$).

\subsection{Equity rule}\label{subsec:Equity}

The heuristic method for solving \gls*{TSA-OPEN} problem using the equity rule modelled in section \ref{sec:Meth} is shown below.  The objective of the \gls*{TSA-OPEN} problem is to minimize the overall deviation of the time slots requested by all the \glspl*{RU} in the system while simultaneously ensuring that the discrepancies between \glspl{RU} do not differ excessively.
The equity criterion is established on the basis of the assigned capacity to each \gls*{RU}. This way, the \gls*{IM} will allocate a time slot to the \gls*{RU} which, at a given time, has received the least number of time slots in relation to its capacity. 
The requests of the \glspl*{RU} are iteratively processed. In each iteration $t$,  the relationship between the allocated time slots and the total capacity of a given \gls*{RU}  (the value  $\phi^t$ ) is recalculated in order to determine which \gls*{RU}'s request is to be served in that iteration. This process continues until all the requests made by the companies are fulfilled. The calculation of $\phi_o^t$ values is shown in Equation~(\ref{eq:phi}). 
\begin{equation}
    \label{eq:phi}
    \phi^t_o = \frac{\sum_{r \in \mathcal{R}}x^t _{or}}{k_o},
\end{equation}
\noindent where $x^t _{or}$ represents the value of the variable $x$ at iteration $t$. After computing the $\phi_o^t$ values for each  \gls{RU} $o$ then \gls{IM} process  the first request from the \gls*{RU} with the lowest value of $\phi_o^t$. If the time slot requested is free, it is directly allocated to the \gls*{RU}. If not, it is re-timed.  Algorithm \ref{code:equity} shows the pseudocode of the heuristic algorithm to solve time slot allocation based on equity rule criterion.

\begin{algorithm}
\begin{algorithmic}[1]
\Require $\mathcal{O}, \mathcal{R}, k, y_1,y_2,...,y_{|\mathcal{O}|}$
\Ensure  $x$
\Function{\gls*{TSA-OPEN} using equity rule}{}
\State -- Initialise $\phi$ values to infinity -- 
\State $\phi = []$
\For{$o=1:|\mathcal{O}|$}
\State $\phi[o] = +\infty$
\EndFor
\State -- Check \glspl*{RU} requests -- 
\State $y = [y_1,y_2,...,y_{|\mathcal{O}|}]$
\State -- Computes the total number of requests received  -- 
\State $n_y = \sum_{o\in{\cal O}} \sum_{r\in{\cal R}} y_{or}$
\State -- Iterative request processing  -- 
\For{$t=1:n_y$} 
\State -- Recalculate $\phi_o$ values  -- 
\State $   \phi_o = \frac{\sum_{r \in \mathcal{R}}x _{or}}{k_o}$
\State $o' = \underset{o \in {\cal O^*}}{\hbox{Arg min}} \{\phi_{o}\}$; where ${\cal O}^*$  is the set of  RU's with time slots that need to be allocated
\State Let $r$ be the next time slot of \gls{RU} $o'$ to be processed. 
\If{$r$ request can be satisfied}
\State $x_{o'r} = 1$
\Else
\State $h^{r'}_{o'r} =1$ where $r'$ is the closet available time slot of the requested $r$.
\State -- Allocate the re-timing time slot -- 
\State $x_{o'r'} =1$ 
\EndIf
\EndFor
\EndFunction
\end{algorithmic}
\caption{Pseudocode for solving heuristically \gls*{TSA-OPEN} problem using an equity rule}\label{code:equity}
\end{algorithm}

In this algorithm, lines $3-6$ defines a $\phi$ vector which contains the initial $\phi$ values for all the \glspl{RU} in the market. Initially, these values are equals to $\infty$. Then, lines $8-10$ allow to obtain the total number of requests issued to the \gls{IM} by all the \glspl{RU}. A for loop in line $12$ goes through all the requests, processing them one by one. In each iteration, $\phi$ values are computed for each \gls{RU}. It will allow to determine the \gls{RU} whose request is to be addressed in each iteration. Once the company to be served is identified, if the time slot requested is free it will be directly allocated (line $18$), if not, the closest available time slot will be assigned (line $22$). 

\section{Numerical Experiments}
\label{sec:NumExp}

This section presents computational experiments using the proposed \gls{TSA-OPEN} approach, applying both the priority and the equity rule criteria. The results provided in this section will allow us to answer the following three \glspl{RQ}:
\begin{itemize}
    \item[\bf RQ1.] How does the use of the two rules proposed in the \gls{IM} affect the equilibrium situation?
    \item[\bf RQ2.] How does the use of heuristic methods instead of exact methods impact the equilibrium situation?
    \item[\bf RQ3.] Are the conclusions still valid if we had used the exact methods to calculate the equilibrium situation?
\end{itemize}

  For these purposes, Section 5.1 describes the considered Spanish railway corridor between Madrid and Barcelona. In section \ref{sec:RQ1}, the impact of the two allocation rules is analysed with respect to the profit of each \gls{RU} (RQ1). Later, section \ref{sec:RQ2} shows the performance of heuristic algorithms against the exact models (RQ2). Finally, section \ref{sec:RQ3} discusses if the conclusions obtained will still be valid if the experiments were carried out using the exact algorithms (RQ3).

%This section describes the computational experiments carried out to demonstrate the performance of the \gls{TSA-OPEN} models under the two rules proposed: priority rule and equity rule. 
%First, the real-life setting at the Spanish railways is described. %It is focused on the Spanish passenger railway market. 
%Then, the two \gls{TSA-OPEN} models are solved using the two greedy algorithms developed in section \ref{sec:Solapp} as well as the  exact algorithms. The results provided by exact and heuristic algorithms allow us to: % define a dual analysis or analysis from two points of view: 
%first, to compare the exact and heuristic methods in the resolution of \gls{TSA-OPEN} problem, and second, to evaluate the performance of solving the problem using priority or equity rule. 
%XX

\subsection{Experimental settings: the Spanish railway market}

\begin{figure}[!h]
    \centering
    \includegraphics[width=0.8\textwidth]{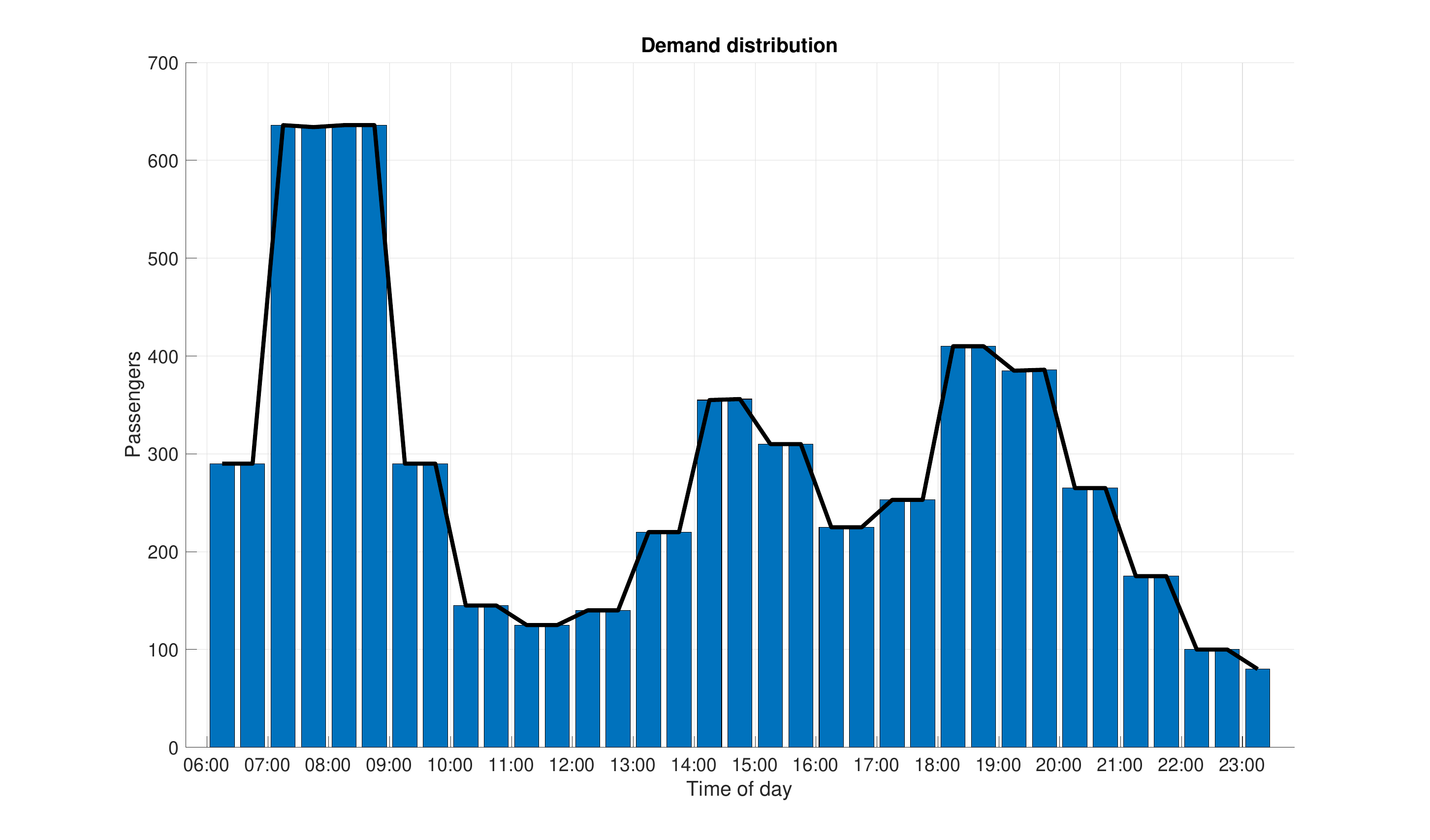}
    \caption{Demand distribution}
    \label{fig:Demand_Distribution}
\end{figure}

The experimental scenario corresponds to the Spanish passenger railway market. Specifically, the Madrid-Barcelona high-speed corridor has been considered. For this purpose, two origin-destination (O-D) pairs have been established: $\omega_1=(\text{Madrid}, \text{Barcelona})$ and $\omega_2=(\text{Barcelona}, \text{Madrid})$. %Regarding the demand model, we consider the following approach: if a company operates in a specific time slot $r$, it captures the entire demand assigned to that time slot.
The demand varies throughout the day according to the distribution given in \cite{AlGa13}, which has been considered for the two pairs $\omega_1$ and $\omega_2$. Figure~\ref{fig:Demand_Distribution} shows the number of passengers travelling in each time slot according to that distribution. We consider that if a company operates in a specific time slot $r$, it captures the entire demand assigned to that time slot.\footnote{Given the levels of demand, this is a reasonable assumption.} 

In the Spanish railway market, there are three \glspl*{RU}: RENFE (the incumbent operator), OUIGO and IRYO, and referred as to $RU_1$, $RU_2$ and $RU_3$ and they are denoted respectively as $o_1$, $o_2$, and $o_3$. Regarding planning time, it is considered one day of operation, starting at 06:15 and ending at 23:15, with a time slot every half hour. Therefore, a total of $35$ time slots in 30-minute intervals are available for each of the origin-destination pairs. In order to study the impact of the allocation policy adopted by the \gls{IM} on the competition between \glspl{RU},  the following assumptions are made. First, all operators have allocated the same capacity, denoted as  $k_o=25\%$, allowing them a maximum capacity of operating 8 time slot per direction.  This is to consider all RUs equally and the results obtained are solely the consequence of the actions of \gls{IM}.  Second,  all \glspl*{RU} employ identical rolling stock and ticket prices. Otherwise, the results obtained could be the result of the policy price or rolling stock difference, rather than the result of the \gls{IM}'s policy. 
%In this work, the same capacity $k_o=(0.25,0.25,0.25)$ has been considered for all \glspl{RU} instead of the real capacity $k_o=(0.65, 0.30, 0.05)$ in order to all \glspl{RU} are indistinguishable and the results obtained are the consequence of the actions of \gls{IM}.

A comprehensive cost framework has been adopted for this analysis. Thus, the ticket price has been set to $70$ euros. The same rolling stock unit can operate several time slots if the time constraints allow it.  The model calculates the minimum number of rolling stock units that a \gls{RU} needs to operate all allocated time slots. Finally, a daily amortisation cost of 11,490 euros has been included for each unit of rolling stock that the \gls{RU} needs to have available to serve its time slots. This cost reflects the depreciation and maintenance expenses associated with the rolling stock units. Furthermore, the direct operational cost of each time slot amounts to $2,950$ euros per slot. This cost factor encompasses various components, such as power consumption and personnel expenses. By considering these cost factors in our analysis, it is intended to provide an  economic evaluation that accounts for both the revenue generation potential through ticket sales and the necessary expenditure required to operate the services efficiently.
%In order to summarise the parameters employed in the computational experiments as well as the assumptions made, 
Table~\ref{tab:settings} summarises the parameters employed in the computational experiments. 

The heuristic algorithms have been coded using the MATLAB programming language and run on a Windows 10 at 4 GHz AMD FX-8370 eight-core processor with 32 GB of RAM. Regarding exact algorithms, CPLEX has been used to solve \gls{TSA-OPEN} problem. The equilibrium model shown in (\ref{eq:model}) has been solved in two stages: (1) the computation of the set of feasible strategies has been carried out according to \cite{Garcia23}, (2) the equilibrium problem is solved in this set by using the algorithm described and implemented in \cite{Chat09}.

\begin{table}[]
\centering
\resizebox{15cm}{!}{
\begin{tabular}{lccl}
\toprule
\multicolumn{4}{c}{\textbf{EXPERIMENTAL SETTINGS}}      \\
\hline
\textbf{Name}                           & \textbf{Notation} & \textbf{Formulation} & \textbf{Value}                                                                                                \\
\hline
O-D pairs & $\mathcal{W}$        & $\mathcal{W}=\{\omega_1,\omega_2\}$              & $\mathcal{W}=\{(\text{Madrid},\text{Barcelona}),(\text{Barcelona},\text{Madrid})\}$ \\
RUs                            & $\mathcal{O}$        &  $\mathcal{O}=\{o_1,o_2,o_3\}$             & $\mathcal{O} = \{RENFE, OUIGO,IRYO\}$                                                                                \\
Time slot                     & $\mathcal{R}$        & $\mathcal{R}=\{r_1,r_2,r_3,..,r_{35}\}$               & $\mathcal{R}=\{06:15,06:45,07:15,...,23:30\}$                                                                                \\
Capacities                     & $k$       & $k=[k_1,k_2,k_3]$              & $k = [25\%,25\%,25\%]$                                                                                \\
Ticket price                   &       $z_{rw}$ &           & 70\euro                                                                                                   \\
Daily amortization cost        &        $C_o$ &           & 11490\euro                                                                                            \\
Cost operational slot         &          $f_{or}$ &           & 2950\euro       \\
\hline
\end{tabular}
}
\caption{Experimental settings set in the numerical experiments}
\label{tab:settings}
\end{table}

\subsection{{\bf RQ1.} \label{sec:RQ1} How does the use of the two rules proposed in the \gls{IM} affect the equilibrium situation?}

In this subsection, the equilibrium problem defined in the equation (\ref{eq:model}) considering both priority and equity rules has been solved. The equilibrium situation is composed by a set of strategies for each \gls{RU}. A pure strategy is defined as a single time slot request (bid). Furthermore, a combined strategy is a set of pure strategies which are played by the \glspl{RU} with a probability value. Table~\ref{tab:TSA-OPEN_solution} shows the obtained equilibrium state for both rules using the heuristic algorithms designed. The column one is associated to the kind of rule, the column two represents the probability to use a combined strategy. Then, columns three and four details the \gls{RU} and the corresponding origin-destination pair. Finally, column $y_o$ shows the request of each \gls{RU} while column $x_o$ represents the time slot finally allocated by the \gls{IM}.

According to Table~\ref{tab:TSA-OPEN_solution}, the solution of the \gls{TSA-OPEN} problem with priority rule consists of a combined strategy for $RU_2$ (composed by two pure strategies, which are played with probabilities $p_{y1}=0.5628$ and $p_{y2}=0.4372$) and one pure strategy for $RU_1$ and $RU_3$ which are employed with probability one.    

Similarly, Table~\ref{tab:TSA-OPEN_solution} also shows the equilibrium situation when solving the \gls{TSA-OPEN} problem considering the equity rule. The equilibrium situation in this case also leads to a combined strategy composed of two pure strategies for $RU_2$. However, while the probabilities in the case of the priority rule for each of the simple strategies were similar, in this case, the probability of $RU_2$ playing the first simple strategy is $p_{y1}=0.0415$ while the probability of playing the second simple strategy is $p_{y2}=0.9585$.

\begin{table}[!h]
\begin{adjustbox}{width=1\textwidth}
\begin{tabular}{llllm{5cm}m{5cm}}
\hline
\textbf{Rule}  & \textbf{Probability  ($p_y$)} & \textbf{\gls{RU}} & $\omega$  & \begin{center}
$y_o$ \end{center} & \begin{center} $x_o$ \end{center}  \\ \hline
\multirow{22}{*}{Priority Rule} 
& \multirow{11}{*}{$p_{y1} = 0.5628$} & \multirow{2}{*}{\gls{RU}1} 
& $\omega_1$ & [7:45, 8:15, 8:45, 9:45, 14:45, 15:15, 18:15, 19:15] &         
               [7:45, 8:15, 8:45, 9:45, 14:45, 15:15, 18:15, 19:15] \\  \cline{4-6}
&            &
& $\omega_2$ & [7:15, 7:45, 8:45, 13:45, 15:45, 18:15, 18:45, 20:15] &
               [7:15, 7:45, 8:45, 13:45, 15:45, 18:15, 18:45, 20:15] \\ \cline{3-6}
&                 & \multirow{2}{*}{\gls{RU}2} 
& $\omega_1$ & [7:45, 8:15, 9:45, 14:45, 15:15, 18:15, 19:15, 20:45] &
               [7:15, 9:15, 10:15, 14:15, 15:45, 18:45, 19:45, 20:45]\\ \cline{4-6}
&                 &                            
& $\omega_2$ & [7:15, 7:45, 8:45, 14:45, 17:45, 18:15, 18:45, 20:15] &
               [6:45, 8:15, 9:15, 14:45, 17:45, 19:15, 19:45, 20:45] \\ \cline{3-6}
&                 & \multirow{2}{*}{\gls{RU}3} 
& $\omega_1$ & [6:45, 7:45, 13:15, 14:45, 15:15, 18:15, 19:15, 20:15] &
               [6:15, 6:45, 13:15, 13:45, 16:15, 17:45, 20:15, 21:15] \\ \cline{4-6}
&            &                            
& $\omega_2$ & [7:15, 7:45, 13:15, 13:45, 15:45, 18:45, 20:15, 20:45] &
               [6:15, 9:45, 13:15, 14:15, 16:15, 17:15, 21:15, 21:45] \\ \cline{2-6}
& \multirow{11}{*}{$p_{y2} = 0.4372$} & \multirow{2}{*}{\gls{RU}1} 
& $\omega_1$ & [7:45, 8:15, 8:45, 9:45, 14:45, 15:15, 18:15, 19:15] &
               [7:45, 8:15, 8:45, 9:45, 14:45, 15:15, 18:15, 19:15] \\ \cline{4-6}
&                 &                            
& $\omega_2$ & [7:15, 7:45, 8:45, 13:45, 15:45, 18:15, 18:45, 20:15] &
               [7:15, 7:45, 8:45, 13:45, 15:45, 18:15, 18:45, 20:15] \\ \cline{3-6}
&                 & \multirow{2}{*}{\gls{RU}2} 
& $\omega_1$ & [7:45, 8:45, 9:45, 14:45, 15:15, 16:45, 18:15, 19:15] &
               [7:15, 9:15, 10:15, 14:15, 15:45, 16:45, 18:45, 19:45] \\ \cline{4-6}
&                 &                            
& $\omega_2$ & [7:15, 7:45, 13:15, 13:45, 15:45, 19:45, 20:15, 20:45] &
               [6:45, 8:15, 13:15, 14:15, 16:15, 19:45, 20:45, 21:15 ]\\ \cline{3-6}
&                 & \multirow{2}{*}{\gls{RU}3} 
& $\omega_1$ & [6:45, 7:45, 13:15, 14:45, 15:15, 18:15, 19:15, 20:15 ] &
               [6:15, 6:45, 13:15, 13:45, 16:15, 17:45, 20:15, 20:45]\\ \cline{4-6}
&                 &                            
& $\omega_2$ & [7:15, 7:45, 13:15, 13:45, 15:45, 18:45, 20:15, 20:45] &
               [6:15, 9:15, 12:45, 14:45, 15:15, 19:15, 21:45, 22:15]\\ \hline
\multirow{22}{*}{Equity Rule} 
& \multirow{11}{*}{$p_{y1} = 0.0415$} & \multirow{2}{*}{\gls{RU}1} 
& $\omega_1$ & [7:45, 8:15, 13:15, 14:45, 15:15, 18:15, 19:15, 19:45] &
               [8:15, 9:15, 12:45, 14:15, 14:45, 19:15, 20:45, 21:45] \\ \cline{4-6}
&                 &
& $\omega_2$ & [7:15, 7:45, 8:45, 13:45, 15:45, 16:45, 18:15, 18:45]  &
               [7:45, 8:45, 9:45, 13:15, 16:45, 17:15, 19:45, 20:15]\\ \cline{3-6}
&                 & \multirow{2}{*}{\gls{RU}2} 
& $\omega_1$ & [6:45, 7:45, 8:15, 15:15, 18:15, 19:15, 19:45, 20:45]  &
               [7:15, 8:45, 9:45, 15:45, 18:45, 19:45, 20:15, 22:15]\\ \cline{4-6}
&                 &                            
& $\omega_2$ & [7:15, 7:45, 8:45, 13:45, 15:45, 18:15, 18:45, 20:15]  &
               [8:15, 9:15, 10:15, 14:45, 16:15, 18:45, 19:15, 20:45]\\ \cline{3-6}
&                 & \multirow{2}{*}{\gls{RU}3} 
& $\omega_1$ & [8:15, 9:45, 13:15, 14:45, 16:45, 18:15, 19:15, 19:45] &
               [7:45, 10:15, 13:45, 15:15, 17:15, 17:45, 18:15, 21:15]\\ \cline{4-6}
&                 &                            
& $\omega_2$ & [7:15, 7:45, 13:45, 15:15, 15:45, 18:15, 18:45, 19:45] &
               [6:45, 7:15, 14:15, 15:15, 15:45, 17:45, 18:15, 21:15] \\ \cline{2-6}
& \multirow{11}{*}{$p_{y2} = 0.9585$} & \multirow{2}{*}{\gls{RU}1} 
& $\omega_1$ & [7:45, 8:15, 13:15, 14:45, 15:15, 18:15, 19:15, 19:45] &
               [8:15, 9:15, 12:45, 15:45, 16:15, 19:15, 20:45, 21:45]\\ \cline{4-6}
&                 &                            
& $\omega_2$ & [7:15, 7:45, 8:45, 13:45, 15:45, 16:45, 18:15, 18:45]  &
               [7:45, 8:45, 9:45, 13:15, 16:45, 17:15, 19:15, 20:15]\\ \cline{3-6}
&                 & \multirow{2}{*}{\gls{RU}2} 
& $\omega_1$ & [6:45, 8:15, 9:15, 13:15, 15:15, 16:45, 18:15, 19:15]  &
               [7:15, 8:45, 9:45, 14:15, 14:45, 17:15, 19:45, 20:15]\\ \cline{4-6}
&                 &                            
& $\omega_2$ & [7:15, 8:45, 13:15, 15:45, 16:45, 18:15, 20:15, 20:45] &
               [8:15, 9:15, 13:45, 16:15, 17:45, 19:45, 20:45, 21:15]\\ \cline{3-6}
&                 & \multirow{2}{*}{\gls{RU}3} 
& $\omega_1$ & [8:15, 9:45, 13:15, 14:45, 16:45, 18:15, 19:15, 19:45] &
               [7:45, 10:15, 13:45, 15:15, 17:45, 18:15, 18:45, 21:15] \\ \cline{4-6}
&                 &                            
& $\omega_2$ & [7:15, 7:45, 13:45, 15:15, 15:45, 18:15, 18:45, 19:45] &
               [6:45, 7:15, 14:15, 15:15, 15:45, 18:15, 18:45, 21:45]  \\ \hline
\end{tabular}
\end{adjustbox}
\caption{Equilibrium solution using the heuristic algorithms and considering the priority and equity rules}\label{tab:TSA-OPEN_solution}
\end{table}

In order to answer RQ1, a two-fold analysis is going to be carried out. It will take into account the time slot allocation and the economic revenue associated with each of the \glspl{RU}. This way, the impact of the two rules in the equilibrium situation will be assessed. 

\begin{figure}[!h]
	\centering
	\subfigure[$\omega_1$]{
		\includegraphics[width=0.47\textwidth]{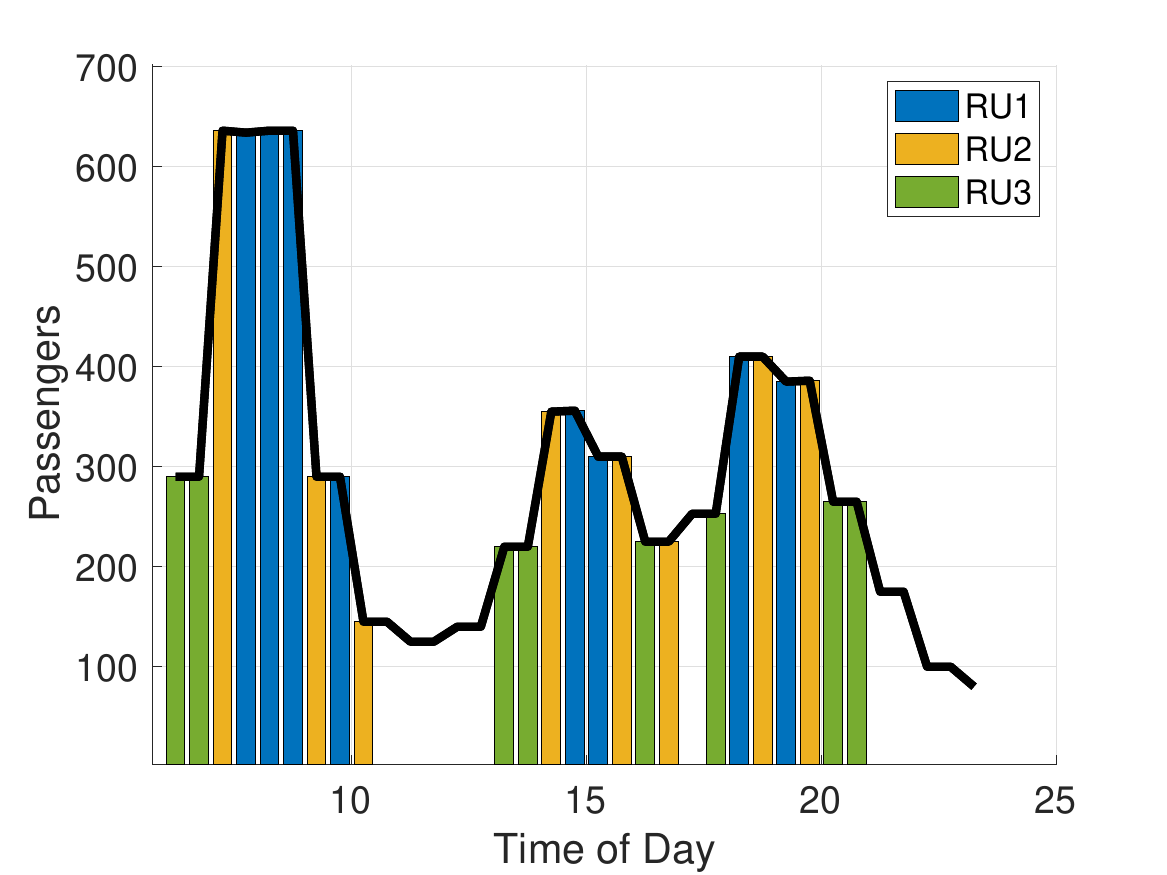}
		\label{fig:HeuPriorO1}
	}
	\subfigure[$\omega_2$]{
		\includegraphics[width=0.47\textwidth]{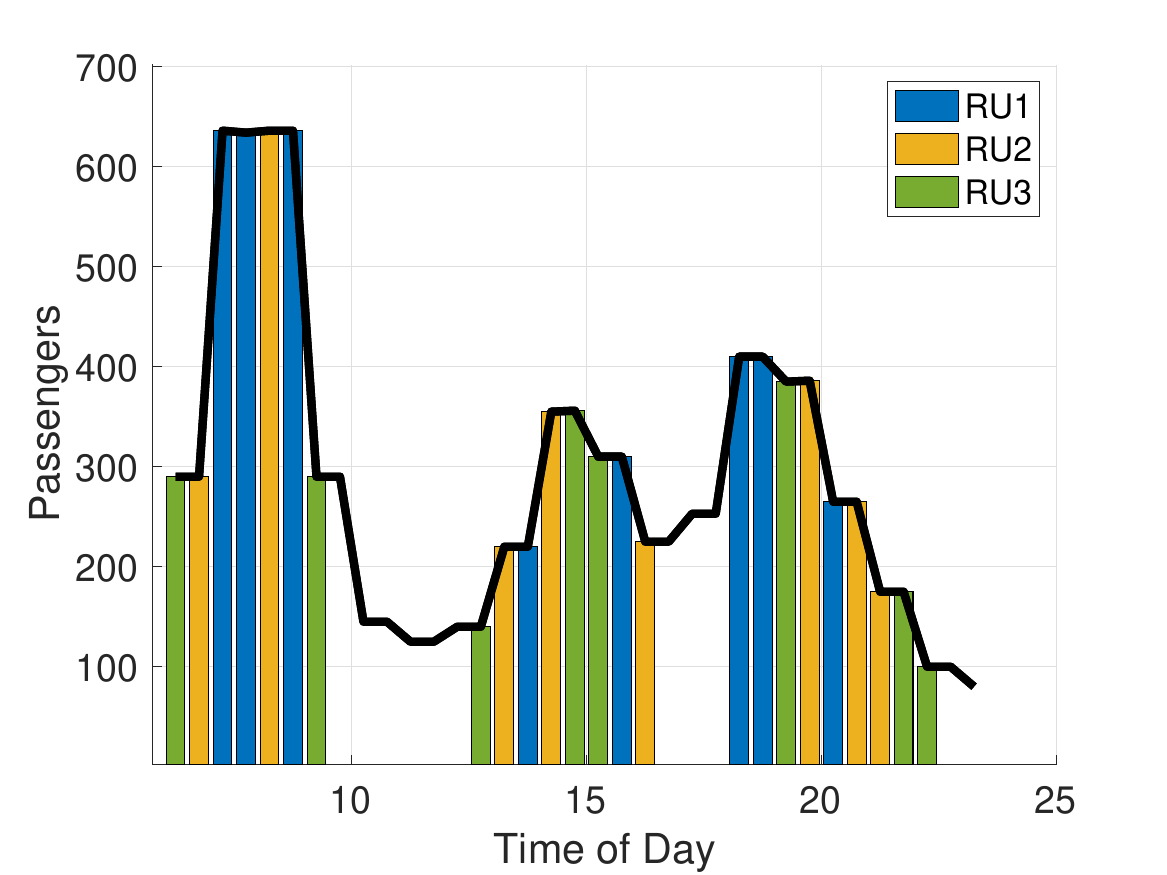}
		\label{fig:HeuPriorO2}
	}
	\caption{Time slot allocation under Priority Rule: Heuristic solution}
	\label{fig:HeuPrior}
\end{figure}

\begin{figure}[!h]
	\centering
	\subfigure[$\omega_1$]{
		\includegraphics[width=0.47\textwidth]{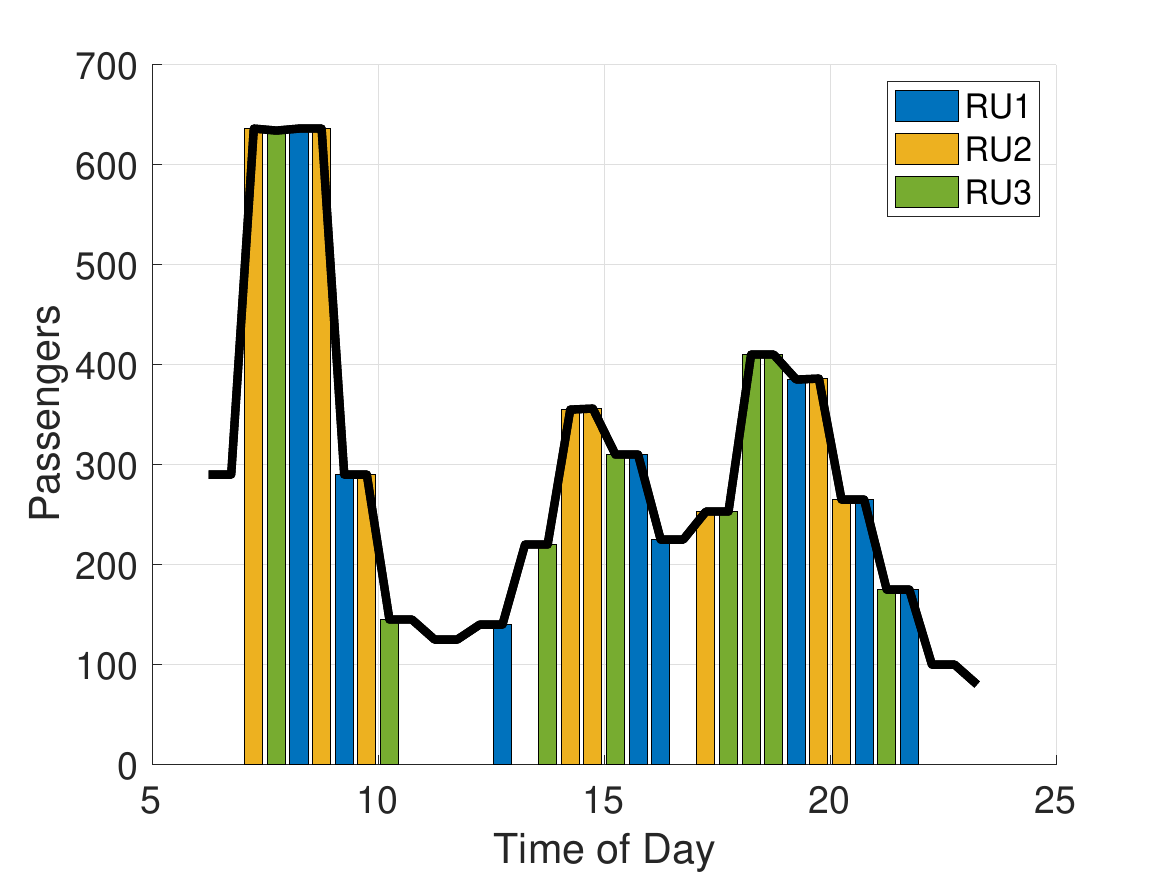}
		\label{fig:HeuEquiO1}
	}
	\subfigure[$\omega_2$]{
		\includegraphics[width=0.47\textwidth]{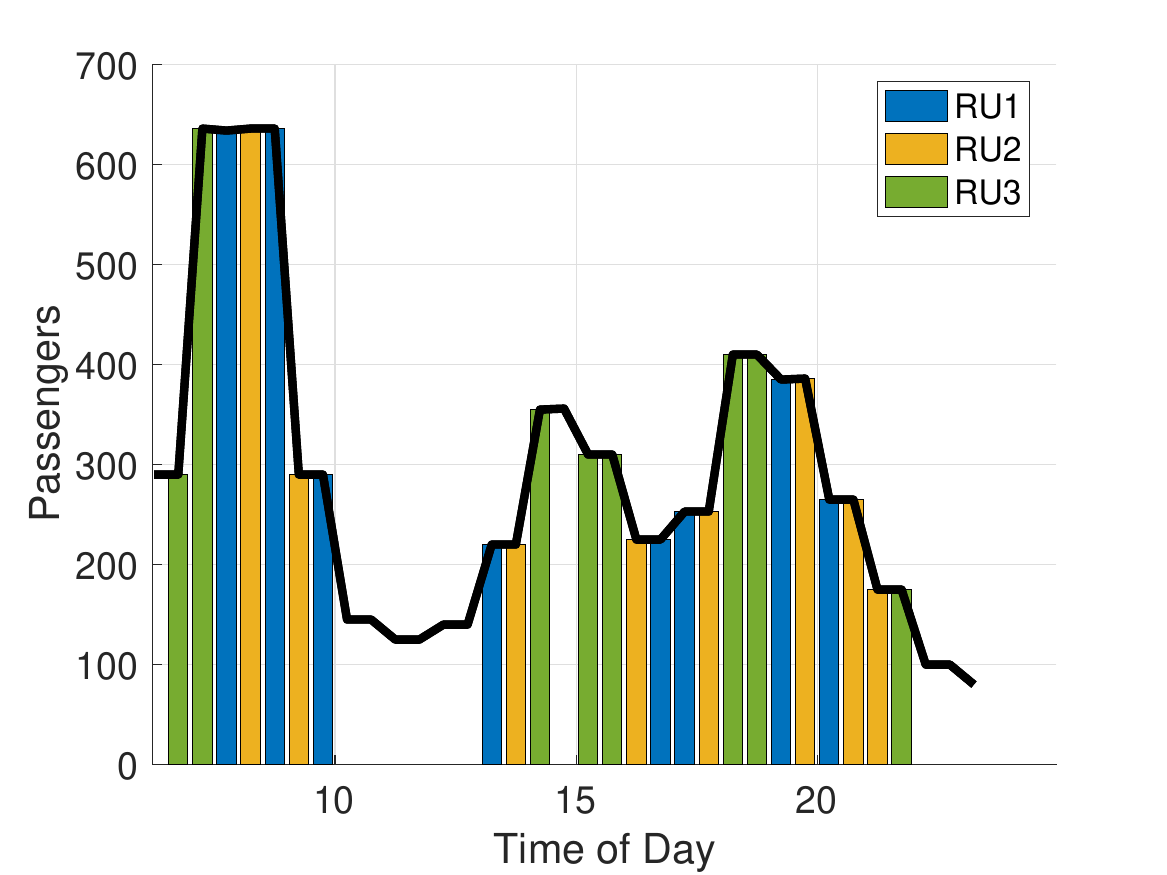}
		\label{fig:HeuEquiO2}
	}
	\caption{Time slot allocation under Equity Rule: Heuristic solution}
	\label{fig:HeuEqui}
\end{figure}

On the one hand, the allocation results will be discussed. Figures~\ref{fig:HeuPrior} and \ref{fig:HeuEqui} show the time slot allocated to the different \glspl{RU} and the demand captured in each one. Concretely, the abscissa axis represents each of the time slots while the ordinate axis represents the number of passengers travelling in each time slot, according to the optimal strategies of the \glspl{RU}\footnote{It is worth noting that the optimal strategy for \glspl*{RU} consists of employing two simple strategies with different probabilities. Here, only one is represented for simplicity.}. Finally, the colour of the bar represents the \gls{RU} to which the time slot has been assigned and on which the passengers are travelling. In addition, each sub-figure displays this information for an origin-destination pair.

Concerning the allocation using priority rule, Figures~\ref{fig:HeuPriorO1} y \ref{fig:HeuPriorO2}, the time slots with the highest demand are allocated to \gls{RU}$_1$, i.e. the one that in the priority rule is served first. Similarly, the time slots with the lowest demand are assigned by the algorithm to the \gls{RU} with the lowest priority. Graphically, both figures show how the \gls{IM}'s actions cause the players' actions to be dominated, i.e. $RU_1$ imposes its criterion on $RU_2$ and $RU_3$ and, in turn, $RU_2$ imposes its criterion on $RU_3$.
The priority rule ensures a preference for the \gls{RU}$_1$ that can lead to a market imbalance and ultimately to a monopoly situation. 

Concerning the allocation using equity rule, Figures~\ref{fig:HeuEquiO1} y \ref{fig:HeuEquiO2}, the main difference with respect to the obtained results from the priority rule is that the allocation of the time slots capturing the highest demand are distributed equally among all \glspl{RU}. Thus, it can be seen that for the two origin-destination pairs, the \glspl{RU} in the market manage to obtain the operating rights of slots that capture high demand, just as they also receive the rights of other slots where the number of passengers is not as large.

On the other hand, the economic revenue of the \glspl{RU} is going to be analysed in order to evaluate the equilibrium situation provided by the solution of \gls{TSA-OPEN} using priority and equity rule. Table~\ref{tab:heuristic_economical_results}  shows, for each of the \glspl{RU} in the market, the number of time slots that have been assigned in each origin-destination pair. In this case, as the capacity of each \gls{RU} is 25\%, each of them receives a total of $16$ time slots, eight in each origin-destination pair. In addition, the table also shows the total number of passengers served by the \glspl{RU} in the assigned time slots in each pair, as well as the total number of passengers served. The fact that decimal numbers appear in these data is because the results in the table are aggregated over the combined strategy, so the total number of passengers is the number of passengers served in each simple strategy 
multiplied by the probability of playing each simple strategy. Finally, the table also shows the number of rolling stock units that each \gls{RU} needs to serve the assigned services as well as the total benefit obtained calculated through the equation (\ref{eq:F}). These values are also computed over the combined strategy, so decimal numbers may appear. These values are obtained after applying an algorithm to compute the minimum number of trains which are required to meet planned schedules. 

\begin{table}[!h]
    \centering
\begin{adjustbox}{width=1\textwidth}
\begin{tabular}{lcrrrrrrr}
\multicolumn{1}{c}{$\cal A$}&\bf \glspl{RU} & $\mathbf{OD}_1$ {\bf Pass.} & $|{\cal R}_1|$ &$\mathbf{OD}_2$ {\bf Pass.}& $|{\cal R}_2|$& \bf  Total Passengers & \bf Rolling Stocks & \bf Revenue \\
\hline
\multirow{3}{*}{\bf Priority rule}
&$o_1$ & 3657 & 8 & 3521 & 8 & 7178 & 6 & 330320 \\ 
&$o_2$ & 2779.5 & 8 & 2725.9 & 8 & 5505.4 & 5.6 & 218262 \\ 
&$o_3$ & 1977.4 & 8 & 2010.5 & 8 & 3987.9 & 4.9 & 119945 \\ 
\hline 
\multirow{3}{*}{\bf  Equity rule}
&$o_1$ & 2433.3 & 8 & 2908 & 8 & 5341.4 & 5 & 213244.8 \\ 
&$o_2$ & 3171 & 8 & 2460.9 & 8 & 5631.9 & 5.1 & 232628.4 \\ 
&$o_3$ & 2550.5 & 8 & 2889.5 & 8 & 5440 & 5 & 220146.8 \\ 
\hline 
\end{tabular}
\end{adjustbox}
\caption{Economical results of solving \gls{TSA-OPEN} problem using heuristic algorithms}
\label{tab:heuristic_economical_results}
\end{table}

According to the results of Table~\ref{tab:heuristic_economical_results}, it is possible to check and contrast the results provided by Figures~\ref{fig:HeuPrior} and \ref{fig:HeuEqui}. In the first row, which corresponds to the results of the priority rule,  there is an imbalance in the market in favour of the  $RU_1$, since it captures many more passengers than the rest of the \glspl{RU} and, therefore, has much greater economic benefits than the rest. Therefore, the impact of priority rule over the equilibrium situation is negative, since it leads to a situation of dominance of the \gls{RU}$_1$ which produces an imbalance in the market, causing some \glspl{RU} to go out of business and, ultimately, leading to a monopoly situation in the market.

Concerning the results of the equity rule which are shown in the second row of Table~\ref{tab:heuristic_economical_results}, it is possible to observe a completely different market situation. In this case, the number of passengers captured by the \glspl{RU} does not differ as significantly as in the previous case, which is also reflected in the number of rolling stock units that each \gls{RU} needs (the same in this case) and in the economic benefits of all of them, which is much more balanced. Therefore, it is possible to verify that the equity rule impacts the equilibrium situation positively, guaranteeing that all \glspl{RU} involved obtain similar benefits, which enables competition within the market.

\subsection{{\bf RQ2.}\label{sec:RQ2} How does the use of heuristic methods instead of exact methods impact the equilibrium situation?}

With the purpose of studying the performance of the two heuristic algorithms proposed to solve \gls{TSA-OPEN} problem using priority and equity rule, exact algorithms have also been used to solve both problems. Concretely, \gls{TSA-OPEN} problem was solved using CPLEX. The results obtained by CPLEX in solving the problem using the priority and the equity rule are shown below. 

On the one hand, Figure~\ref{fig:TSA-OPEN Priority} shows the allocation results provided by heuristic and exact algorithms when \gls{TSA-OPEN} problem with priority rule is solved. In this figure, an image matrix composed of three rows and two columns can be observed. Each row corresponds to a different \gls{RU} in the market, while each column corresponds to a different origin-destination pair. Thus, Figures~\ref{fig:Exp1PriRU1W1} and \ref{fig:Exp1PriRU1W2} correspond to $RU_1$ in both directions when applying the priority rule. For each figure in the matrix, a so-called ring diagram shows the requested and allocated slots, where the outer ring with the thickest line represents the requested slots by the \glspl{RU}, the middle ring shows the allocation provided by the heuristic algorithm while the inner ring with the thinnest line corresponds to the allocation obtained by the exact algorithm. 

\begin{figure}[!h]
	\centering
	\subfigure[$RU_1: \omega_1$]{
		\includegraphics[width=0.47\textwidth]{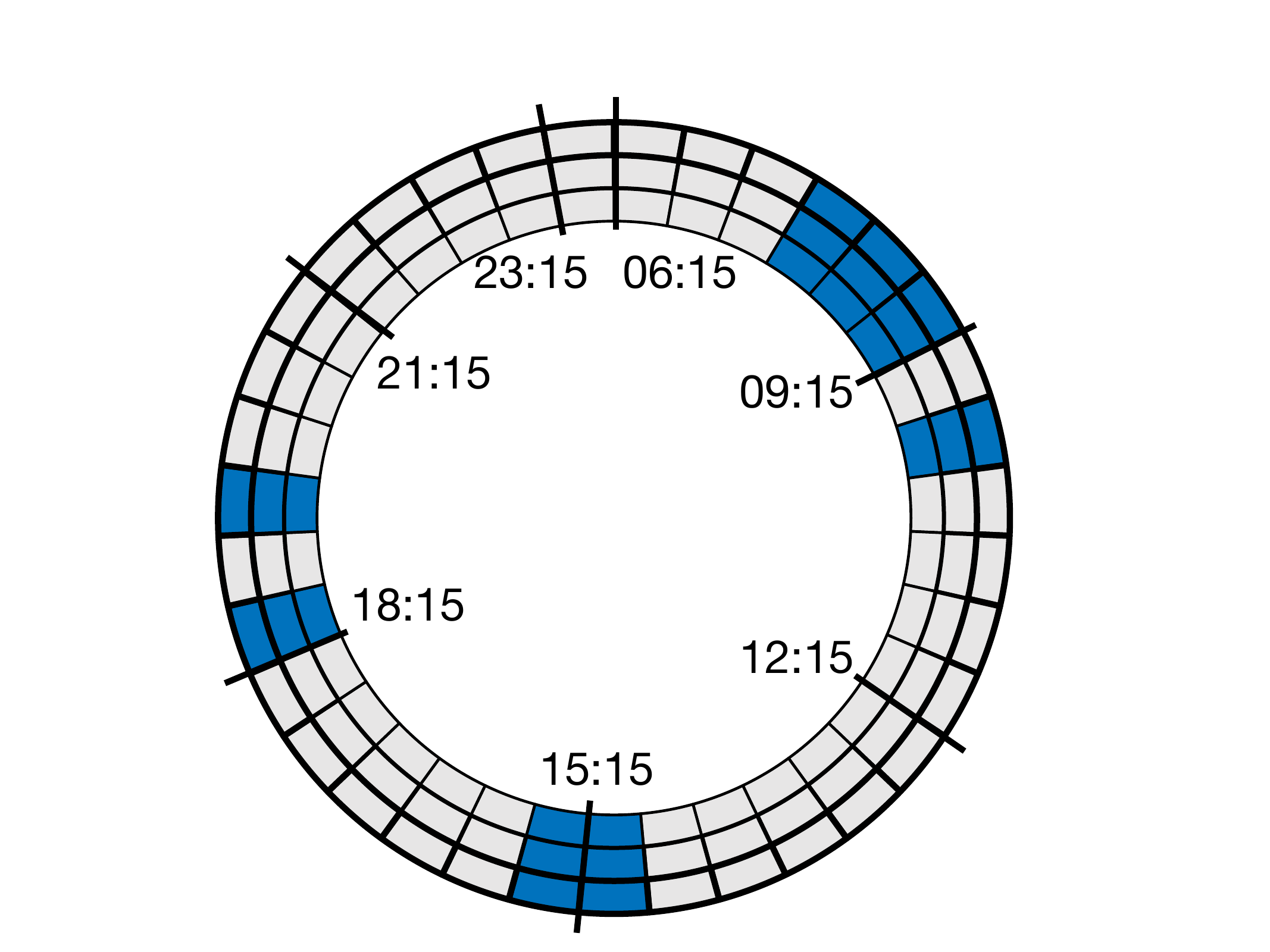}
		\label{fig:Exp1PriRU1W1}
	}
	\subfigure[$RU_1: \omega_2$]{
		\includegraphics[width=0.47\textwidth]{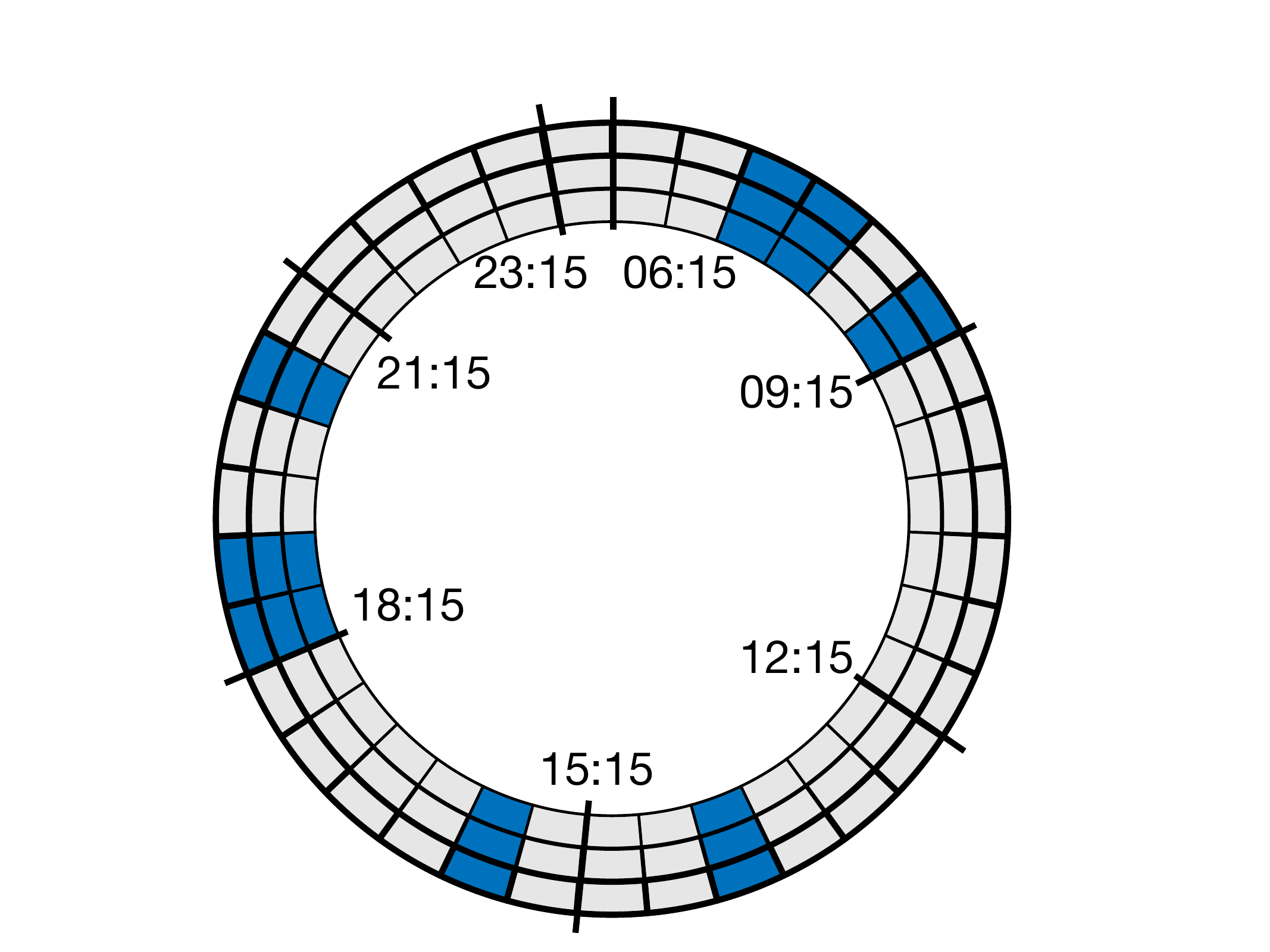}
		\label{fig:Exp1PriRU1W2}
	}
 	\subfigure[$RU_2: \omega_1$]{
		\includegraphics[width=0.47\textwidth]{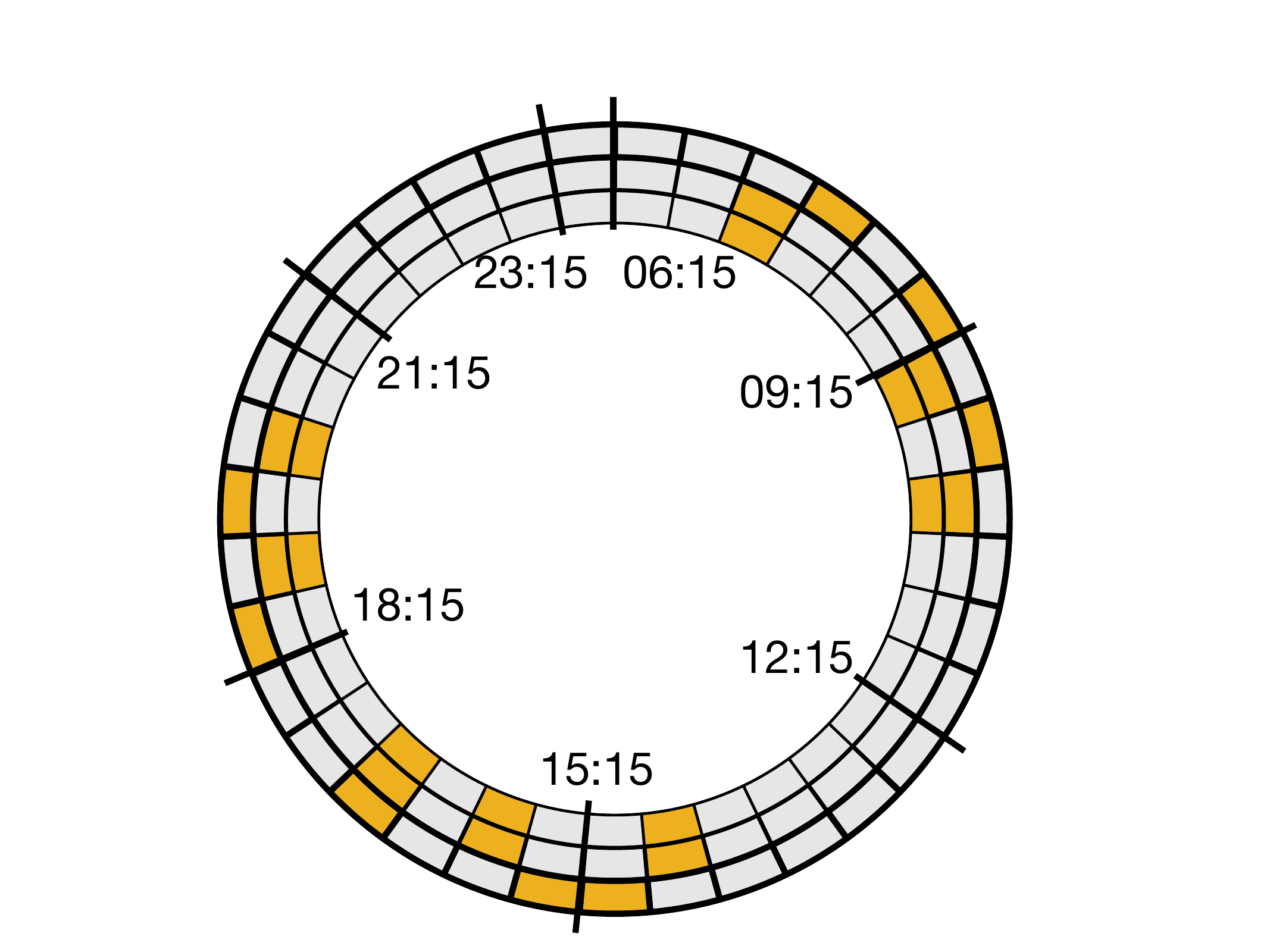}
		\label{fig:Exp1PriRU2W1}
	}
 	\subfigure[$RU_2: \omega_2$]{
		\includegraphics[width=0.47\textwidth]{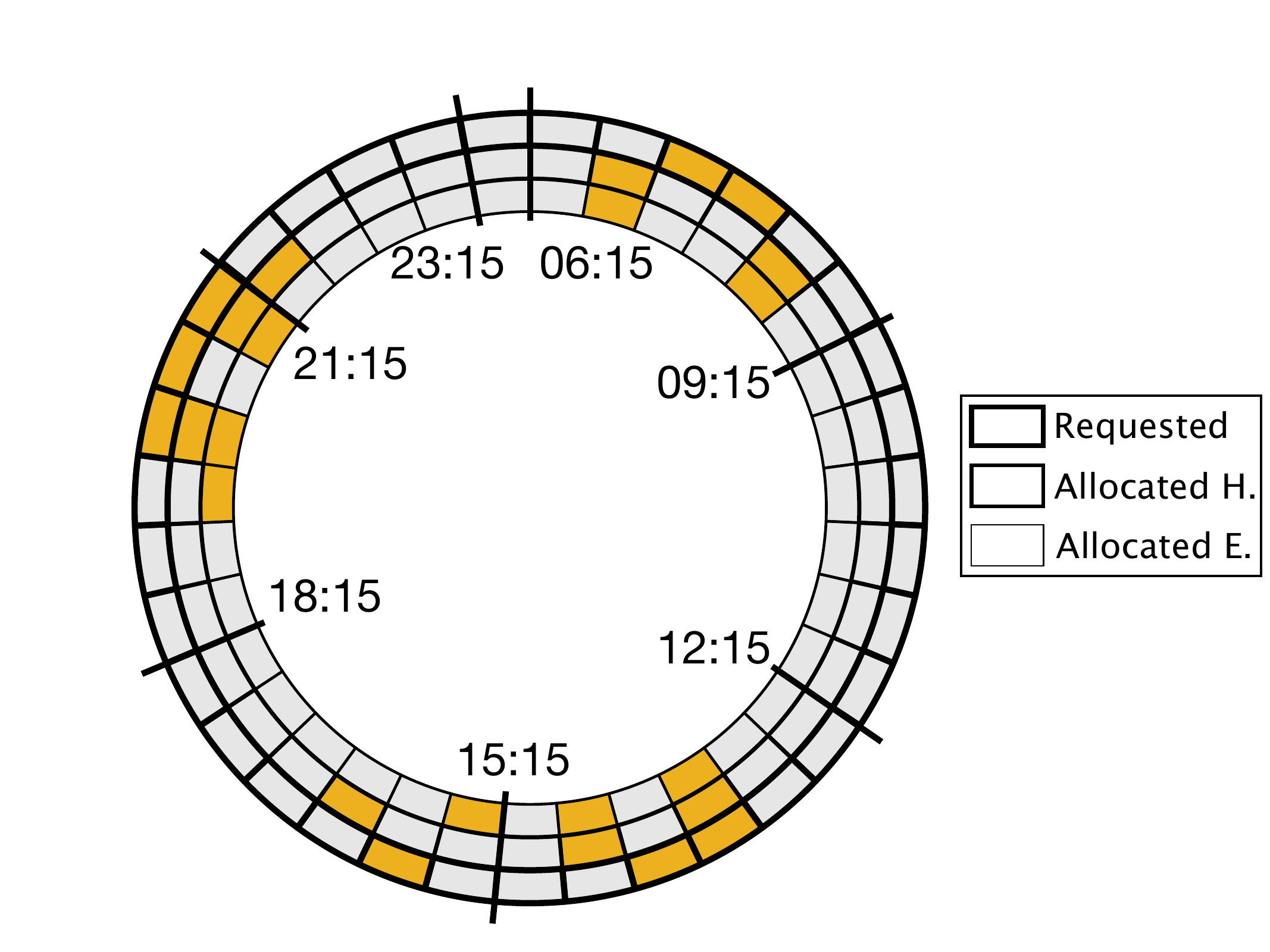}
		\label{fig:Exp1PriRU2W2}
	}
  	\subfigure[$RU_3: \omega_1$]{
		\includegraphics[width=0.47\textwidth]{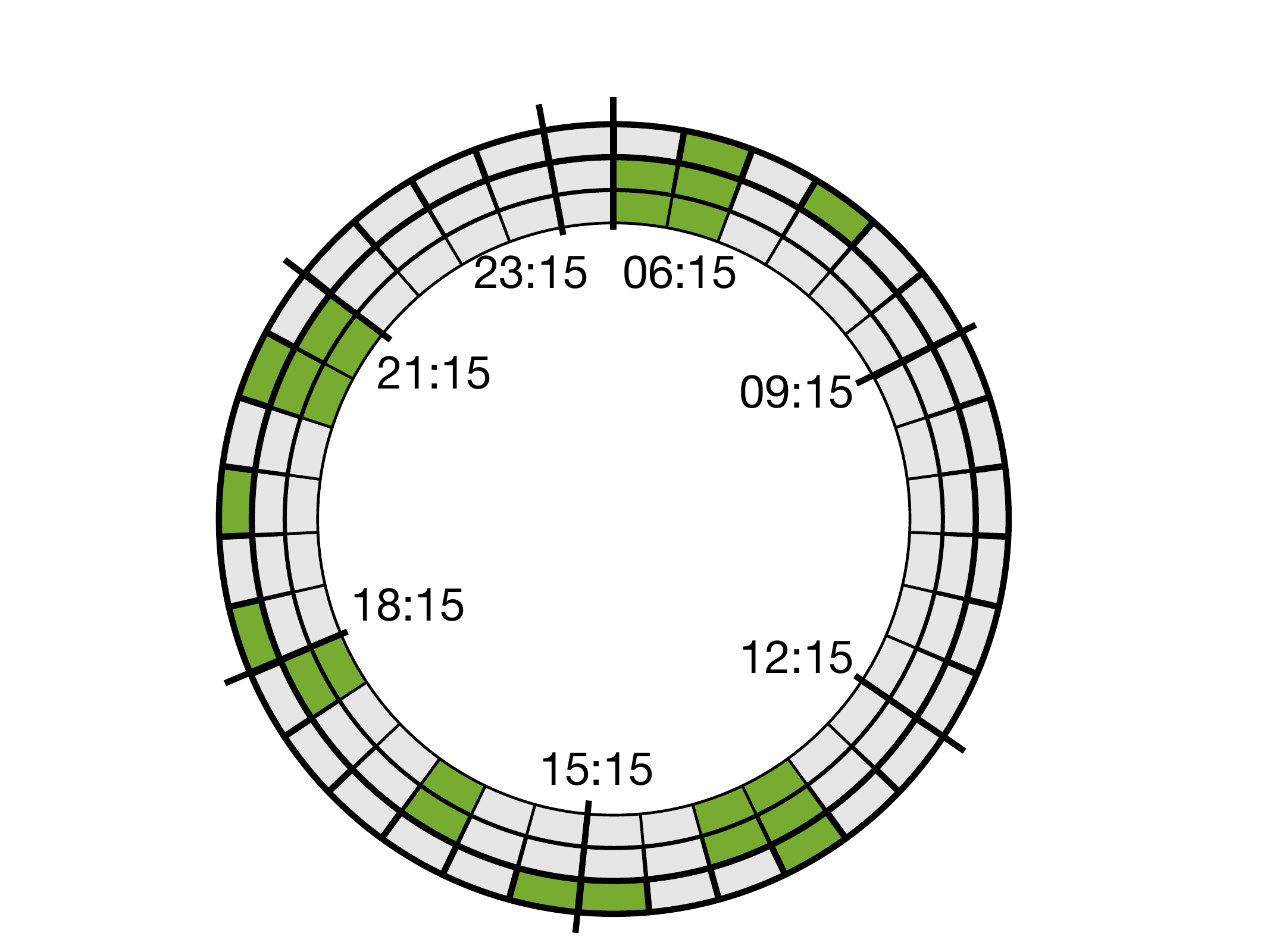}
		\label{fig:Exp1PriRU3W1}
	}
  	\subfigure[$RU_3: \omega_2$]{
		\includegraphics[width=0.47\textwidth]{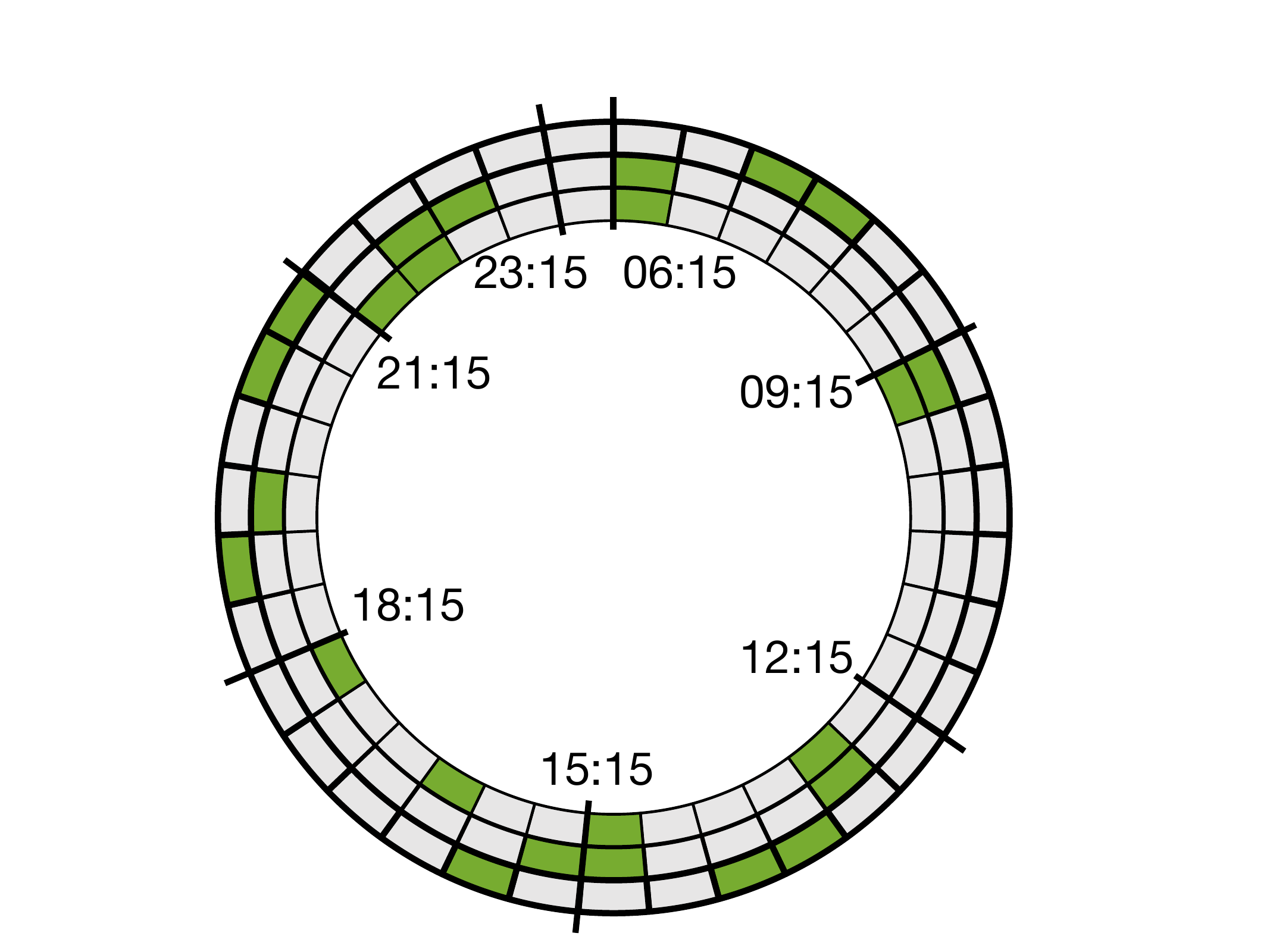}
		\label{fig:Exp1PriRU3W2}
	}
	\caption{\gls{TSA-OPEN} solution under priority rule. Each row represents a \gls{RU} in the market and each column an origin-destination pair. The graphs compare the requested time slot (outer ring) to those one allocated by using heuristic (middle ring) and exact (inner ring) algorithms.}
	\label{fig:TSA-OPEN Priority}
\end{figure}

Analysing Figure~\ref{fig:TSA-OPEN Priority} it is possible to contrast that, for $RU_1$, the allocation provided for both heuristic and exact algorithms is the same in the two origin-destination pairs. Regarding $RU_2$, the allocation for the first pair obtained by both algorithms is also the same. However, the allocation provided for the second pair $\omega_2$ is slightly different in both algorithms, allocating six of the eight time slots equally by both algorithms. Finally, something very similar happens with $RU_3$, where the allocation for $\omega_1$ origin-destination pair is the same using both exact and heuristic algorithms, but in the case of the second pair, only five of the eight time slots are allocated equally by both algorithms. In this test scenario in which all companies have the same capacity, the priorities were assigned to $RU_1$, $RU_2$ and $RU_3$ respectively (note that if priorities were assigned according to the capacity of the \glspl{RU}, the incumbent operator will be favoured). As can be seen in this figure, the coincidence of both algorithms in the solution is an indicator of the quality of the proposed heuristic algorithm.

\begin{figure}[!h]
	\centering
	\subfigure[$RU_1: \omega_1$]{
		\includegraphics[width=0.47\textwidth]{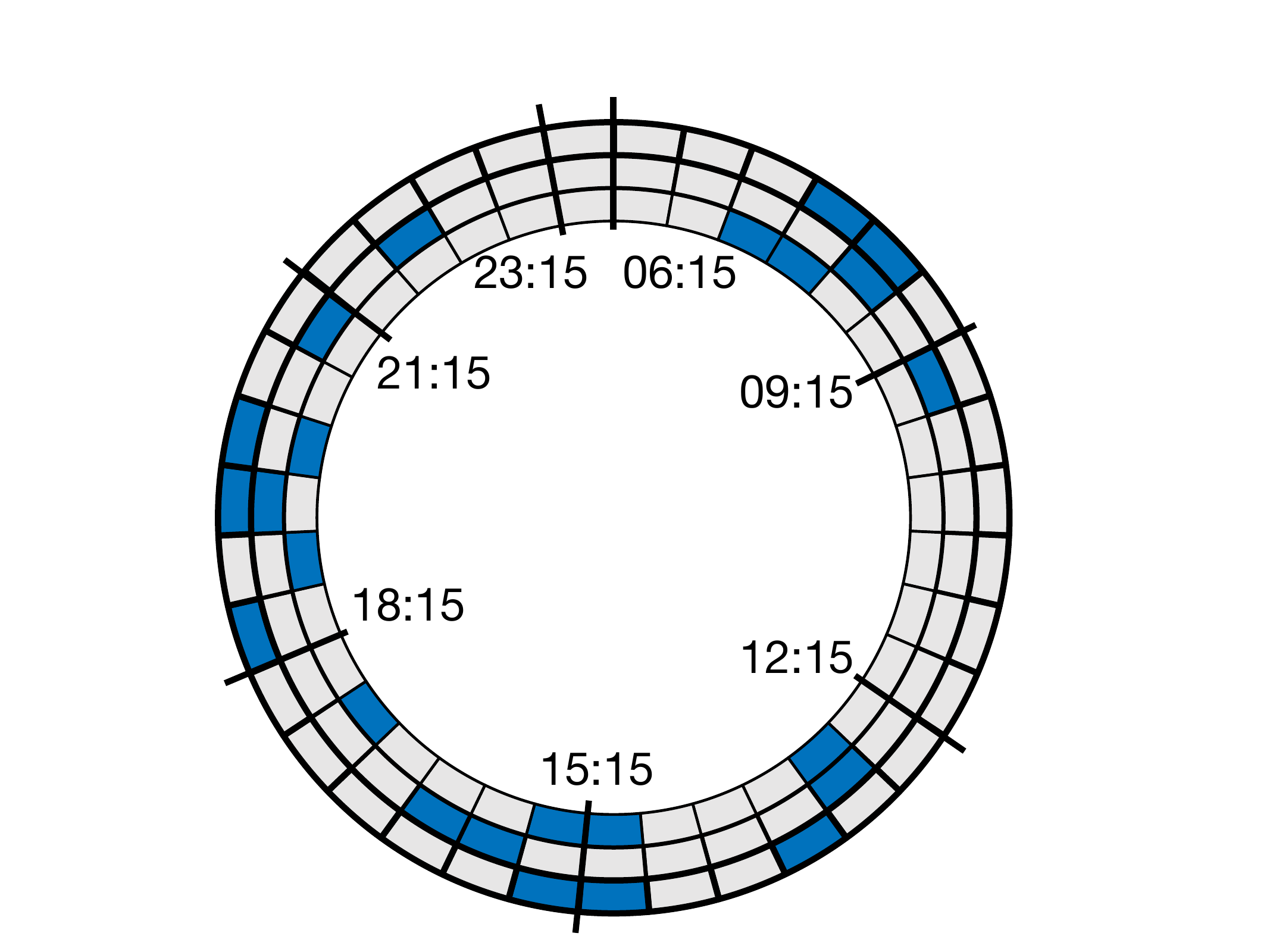}
		\label{fig:Exp1EquiRU1W1}
	}
	\subfigure[$RU_1: \omega_2$]{
		\includegraphics[width=0.47\textwidth]{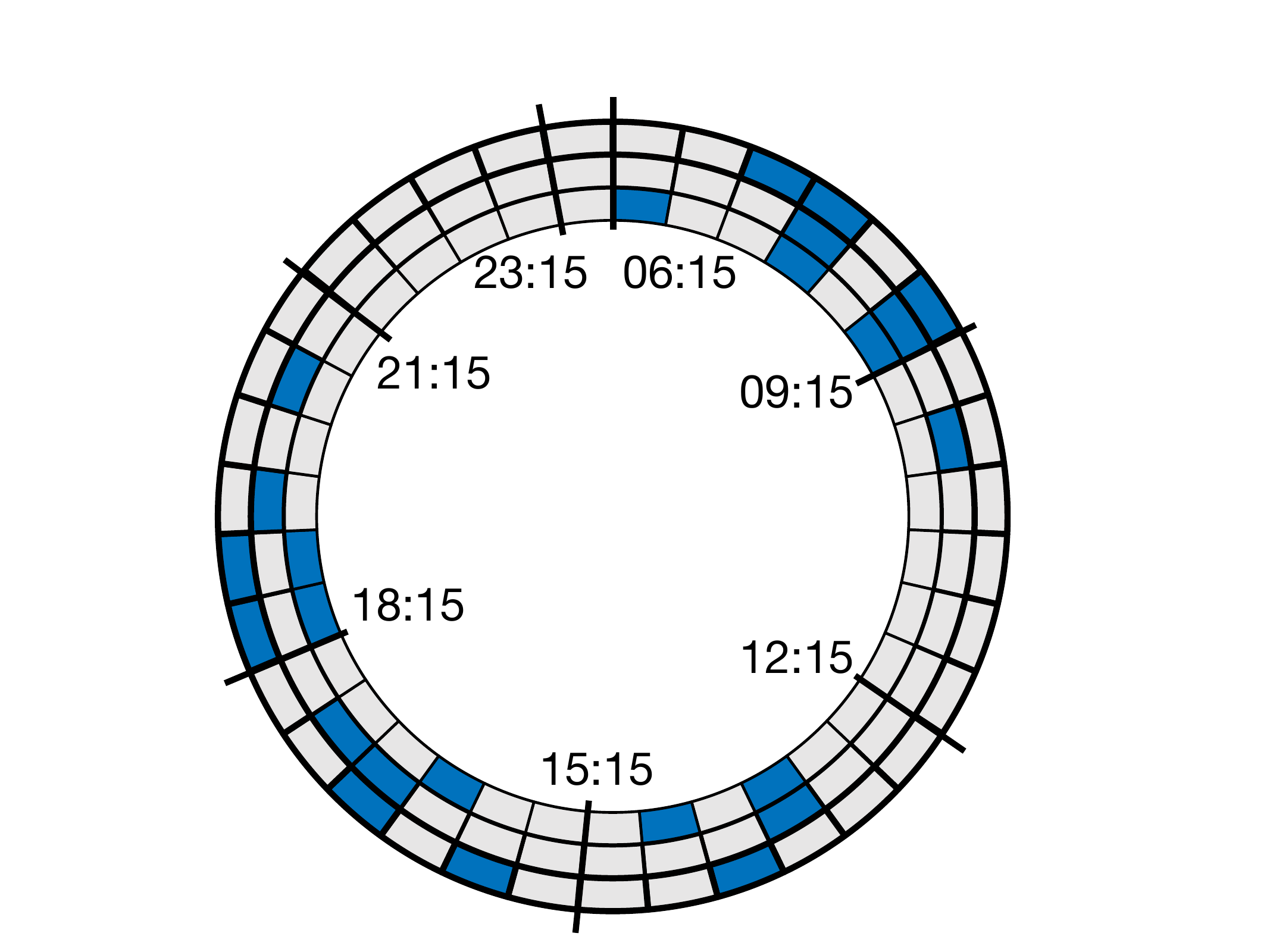}
		\label{fig:Exp1EquiRU1W2}
	}
 	\subfigure[$RU_2: \omega_1$]{
		\includegraphics[width=0.47\textwidth]{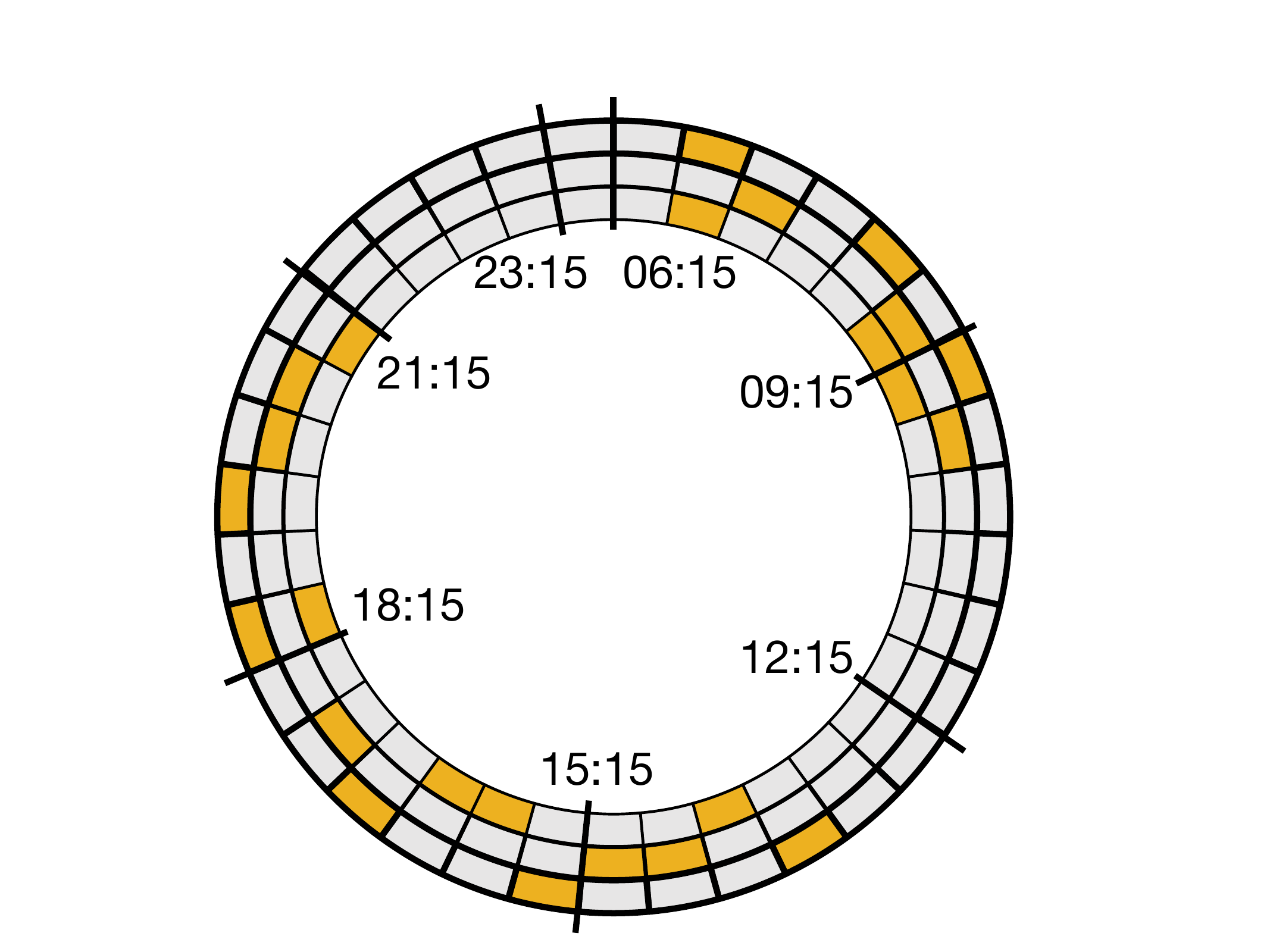}
		\label{fig:Exp1EquiRU2W1}
	}
 	\subfigure[$RU_2: \omega_2$]{
		\includegraphics[width=0.47\textwidth]{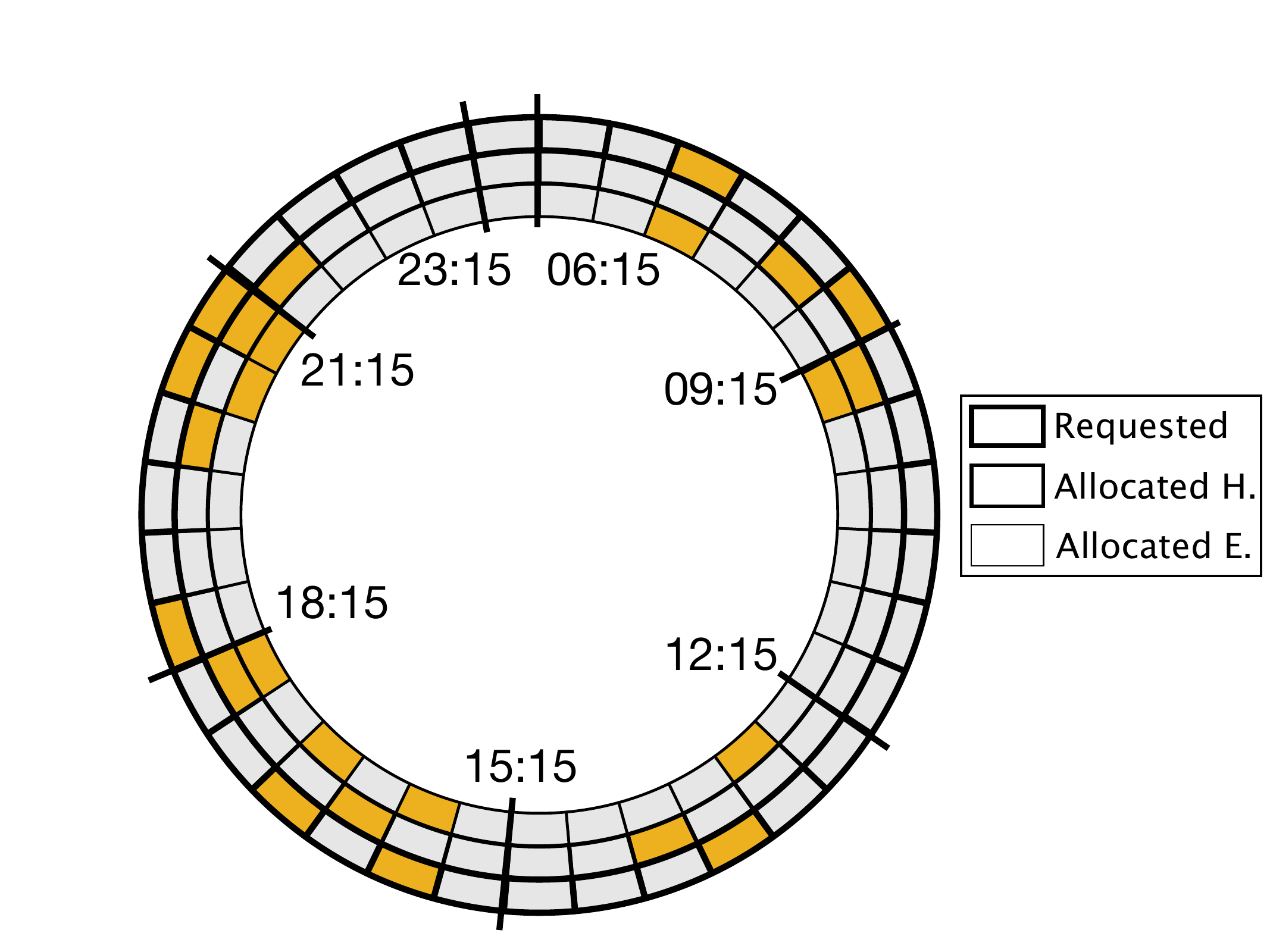}
		\label{fig:Exp1EquiRU2W2}
	}
  	\subfigure[$RU_3: \omega_1$]{
		\includegraphics[width=0.47\textwidth]{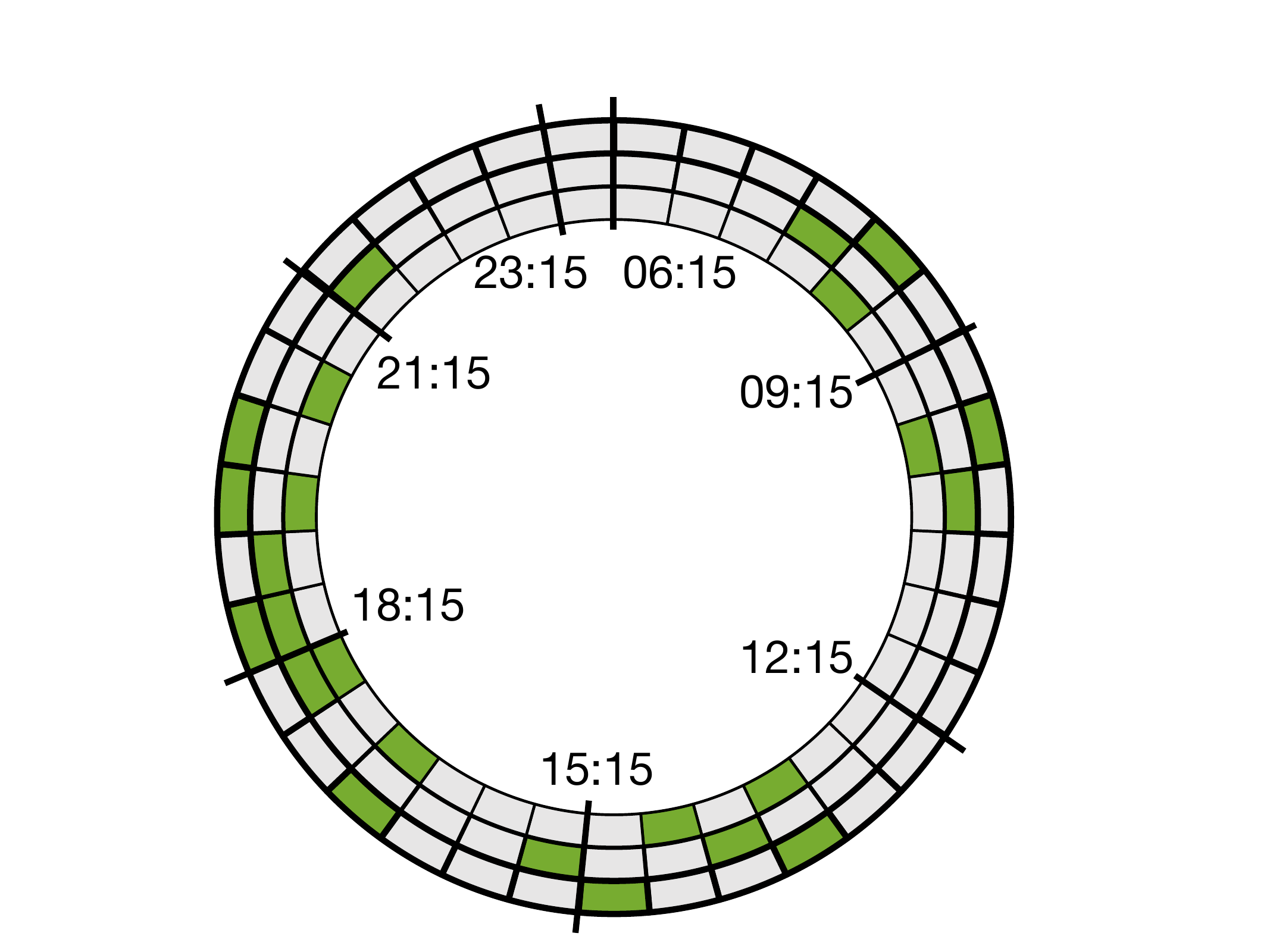}
		\label{fig:Exp1EquiRU3W1}
	}
  	\subfigure[$RU_3: \omega_2$]{
		\includegraphics[width=0.47\textwidth]{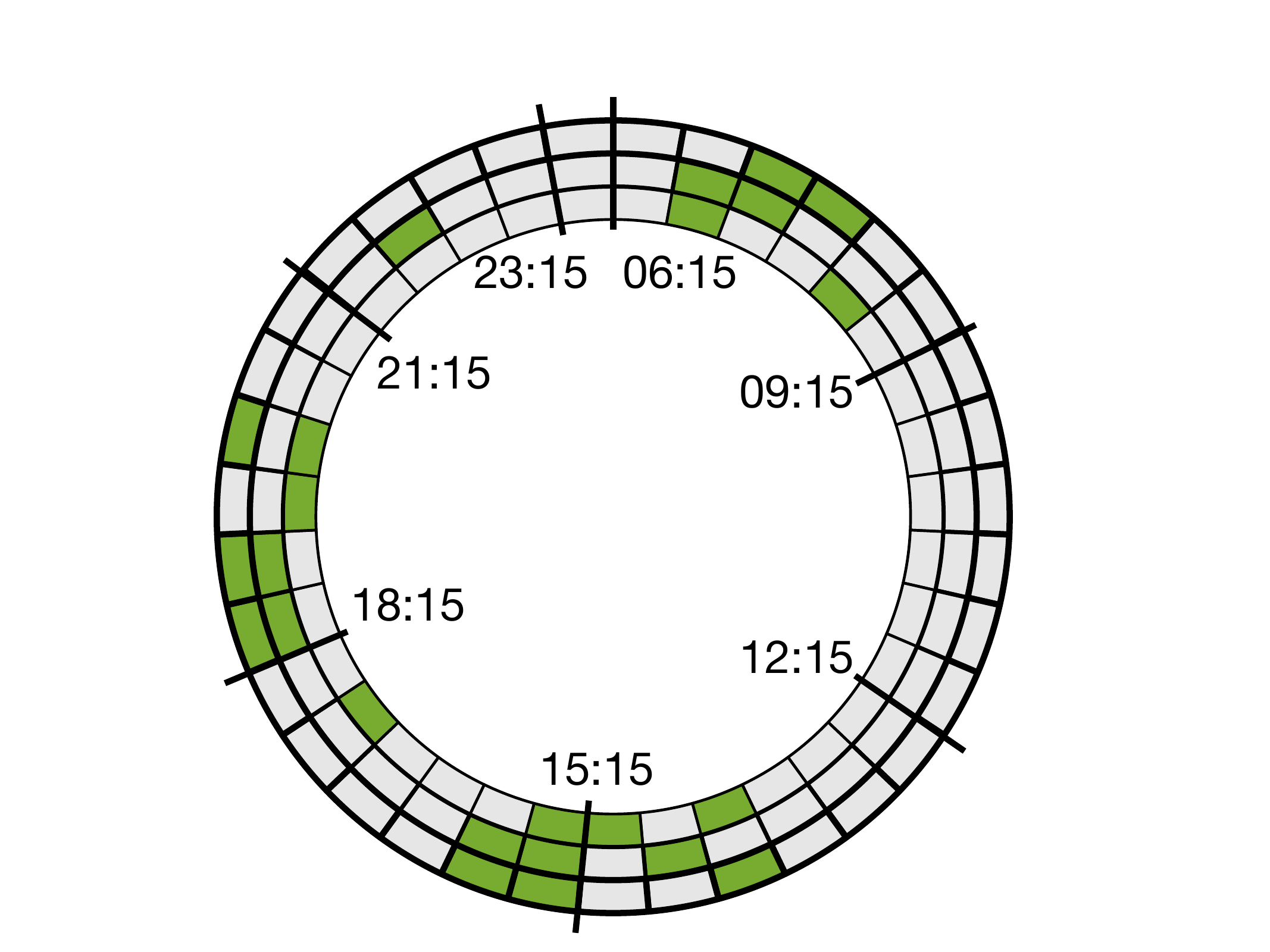}
		\label{fig:Exp1EquiRU3W2}
	}
	\caption{\gls{TSA-OPEN} solution under equity rule. Each row represents a \gls{RU} in the market and each column is an origin-destination pair. The graphs compare the requested time slot (outer ring) to those allocated by using heuristic (middle ring) and exact (inner ring) algorithms.}
	\label{fig:TSA-OPEN Equity}
\end{figure}

On the other hand, Figure~\ref{fig:TSA-OPEN Equity} shows the results of heuristic and exact algorithms when \gls{TSA-OPEN} is solved by means of equity rule. Here, the results are quite different. However, these results are reasonable since, under this approach, the time slots are proportionally distributed according to the \glspl{RU}'s capacities. In this case, the solutions of exact and heuristic approaches are different. Thus, it is possible to check that, for $RU_1$, only four and three time slots respectively are equally assigned by both algorithms for the two origin-destinations pairs. For $RU_2$ in the $\omega_1$ pair, two time slots are equally assigned by heuristic and exact algorithms while in $\omega_2$, there are three time slots. Finally, regarding $RU_3$, only one time slot is assigned by both heuristic and exact algorithms for the first pair, while in the second origin-destination pair, three time slots are equally assigned by both algorithms. 

In order to summarize the results of Figures~\ref{fig:TSA-OPEN Priority} and \ref{fig:TSA-OPEN Equity}, Table~\ref{tab:deviations} shows a study of the deviation of each of the \glspl{RU}. The deviation  for a \gls{RU} is the sum of the time differences between the time slot it requested and the time slot it was finally allocated. Thus, the smaller the deviation, the more the allocation of time slot was in line with the \gls{RU}'s request, while the larger the deviation, the greater the discrepancy between the time slot requested and those finally allocated to the \gls{RU}. This way, Table~\ref{tab:deviations} shows this comparison. It is possible to check the performance of heuristic and exact algorithms is similar when the priority rule is employed, since the deviation for $RU_1$ and $RU_2$ is the same, while the difference in the deviation of $RU_3$ is only half an hour, it means, only one time slot deviation. Thus, the total deviation using the priority rule is around twenty hours. However, in the case of equity rule, the differences are greater. The heuristic approach exhibits an average deviation of approximately ten hours, whereas the exact approach yields half the deviations for all the \glspl{RU}.

\begin{table}[!h]
\begin{tabular}{llllll}
\bf $\mathcal{A}$ & \textbf{Algorithm} & \bf $D_{o1}$ & \bf $D_{o2}$ & \bf $D_{o3}$  & \textbf{Total Deviation} \\ \hline
\multirow{2}{*}{\bf Priority Rule} & Heuristic & 0    & 6h 30m & 13h 30m & 20h              \\
                               & Exact     & 0    & 6h 30m & 13h   & 19h 30m              \\ \hline
\multirow{2}{*}{\bf Equity Rule}   & Heuristic & 15h 30m & 12h  & 8h 30m   & 36h            \\
                               & Exact     & 6h 30m  & 5h   & 5h   & 16h 30m \\ \hline          
\end{tabular}
\caption{Deviation analysis of \gls{TSA-OPEN} solutions}\label{tab:deviations}
\end{table}

The results provided by Table~\ref{tab:deviations} seem to indicate that, when solving the \gls{TSA-OPEN} problem with priority rule, it is indifferent to use heuristic and exact algorithms. However, when attempting to address the \gls{TSA-OPEN} problem using the equity rule, exact algorithms ensure an optimal allocation of time slots with respect to the disparity between the requests made by the different carriers. To validate this observation from the table and considering that temporal disparity differs from the economic outcomes each of the \glspl{RU} would achieve, we will reevaluate the equilibrium strategy using exact methods. The outcomes are presented in Table~\ref{tab:exact_economical_results}.

From Table~\ref{tab:exact_economical_results}, it can be seen that in terms of the allocation results with priority rule, the economic benefits obtained are very similar (slightly higher) than those obtained by the heuristic algorithms, both in terms of the number of passengers travelling on each company and the benefit obtained by each of them. These results, therefore, are in line with those obtained in Table~\ref{tab:deviations}, which showed how the discrepancy obtained for each of the companies by the heuristic and exact algorithms was similar, although slightly better than that obtained by the exact algorithms. 

On the other hand, the results obtained when the exact algorithms are used to solve the \gls{TSA-OPEN} problem with the equity rule are striking. In this particular case, both the number of passengers and the economic benefit obtained by each of the \glspl{RU} differ significantly from those obtained by the heuristic algorithms. These economic differences are due to the discrepancy obtained by the heuristic algorithms and the exact algorithms (see Table~\ref{tab:deviations}). However, the results obtained by the exact algorithms, despite guaranteeing a smaller discrepancy between the \glspl{RU}' requests and the time slots finally allocated,  maintain a significant imbalance in terms of the number of passengers travelling on each \gls{RU} and the profits of each \gls{RU}, showing a non-equilibrium solution with respect to the exact algorithms.

Despite this negative outcome regarding the quality of the heuristic using the equity rule, there is a significant highlight: when addressing the \gls{TSA-OPEN} problem with the equity rule, it becomes evident that the \glspl{RU} can learn how the \gls{IM} operates and devise a strategy that leads to an equilibrium ensuring equal or similar profits for all of them. Therefore, if the \gls{IM} sets a rule that is transparent and does not lead to a dominant strategy of one over the others, the \glspl{RU} will come to play a strategy that allows them to reach an equilibrium situation and perceive similar profits. In other words, what is important is not only that the discrepancy between requested and allocated time slots is small, but that all \glspl{RU} in the market are treated equally by the \gls{IM}. 

\begin{table}[]
    \centering
\begin{adjustbox}{width=1\textwidth}
\begin{tabular}{lcrrrrrrr}
\multicolumn{1}{c}{$\cal A$}&\bf TOCs & $\mathbf{OD}_1$ & $|{\cal R}_1|$ &$\mathbf{OD}_2$& $|{\cal R}_2|$& \bf  Passengers & \bf Rolling Stocks & \bf Revenue \\
\hline
\multirow{3}{*}{\bf Priority rule}
&$o_1$ & 3657 & 8 & 3521 & 8 & 7178 & 6 & 330320 \\ 
&$o_2$ & 2779.5 & 8 & 2780 & 8 & 5559.5 & 5 & 228517.7 \\ 
&$o_3$ & 2021.2 & 8 & 2071.2 & 8 & 4092.4 & 5.1 & 124374.7 \\ 
\hline 
\multirow{3}{*}{\bf Equity rule}
&$o_1$ & 3134.8 & 8 & 3178.2 & 8 & 6313 & 5 & 281260.6 \\ 
&$o_2$ & 2649.9 & 8 & 2411.8 & 8 & 5061.7 & 5 & 193192.5 \\ 
&$o_3$ & 2614.7 & 8 & 2816.2 & 8 & 5430.9 & 5 & 219512.1 \\ 
\hline 
\end{tabular}
\end{adjustbox}
\caption{Economical results of solving \gls{TSA-OPEN} problem using exact algorithms }
\label{tab:exact_economical_results}
\end{table}

\subsection{{\bf RQ3.}\label{sec:RQ3} Are the conclusions still valid if we had used the exact methods to calculate the equilibrium situation?}
To answer this question, it is necessary to start from the results obtained in the previous section. As indicated in Table~\ref{tab:exact_economical_results} compared to Table~\ref{tab:heuristic_economical_results}, the results obtained by the heuristic and exact algorithms for the resolution of the \gls{TSA-OPEN} with priority rule are similar in terms of discrepancy, number of passengers captured and economic benefits obtained by the \glspl{RU}. In both cases, it can be observed that the priority rule imposes a dominance of the  \gls{RU}$_1$ over the rest in the market, which jeopardises the situation of liberalisation and competition. Therefore, in this case, the conclusions are the same whether exact or heuristic algorithms are used to solve the \gls{TSA-OPEN} problem.

However, when addressing the \gls{TSA-OPEN} problem with an equity rule, the conclusions differ based on whether heuristic or exact algorithms are employed for problem-solving. As demonstrated in the previous section, the use of exact algorithms ensures an optimal alignment between the time slots requested by the \glspl{RU} and those ultimately assigned. However, this criterion has not been employed as the criterion $x={\cal A}(y)$ in the equilibrium calculation (see Equations~(\ref{eq:equilibrio1})-(\ref{eq:equilibrio2})), and as a result, it does not lead to a balance in the benefits received by the companies. To obtain a definitive answer for {\bf R3} in the case of the equity rule, equilibrium calculation using exact methods for the allocation is required. The challenge lies in the necessity to solve a large number of linear programming problems, which is the motivation behind utilizing heuristic methods.

One can postulate the response to {\bf R3} for the equity rule using exact methods based on the insights gained from heuristic approaches. In alignment with the results presented in the previous section, where all companies achieve comparable revenues despite a substantial divergence between the requested and allocated time slots, it becomes apparent that these companies familiarize themselves with the operations of the \gls{IM} and formulate strategies to attain an equilibrium in which they all receive equal benefits.

The conjecture to assert that the \glspl{RU} will reach  an {\sl equitable equilibrium}, i.e.  indistinguishable companies achieve the same revenue,  is as follows: If the method utilized, denoted as ${\cal A}(y)$, were universally known among all \glspl{RU}, and furthermore, if ${\cal A}$ possessed the property that swapping the requests $y_o$ of two \glspl{RU} results in the \gls{IM} reciprocally swapping the time slot assignments, then the equilibrium would indeed be equitable, and the role of the \gls{IM} would be neutral in this context.

\section{Conclusions and further works}
\label{sec:Conclusions}

The European passenger railway markets are progressing steadily towards the creation of a common and unified railway area, where multiple \glspl{RU} compete with each other for operating rights in time slots, which means a greater supply for passengers and an opportunity to improve rail services. This evolution, which is marked by the competitive aspect of the liberalised rail markets, is leading to changes in railway planning, operation and management, which require new models and algorithms for their efficient resolution.  

This paper is a novel research work about the modelling of open passenger railway systems. The first contribution is the modelling of \gls{TSA-OPEN} problem, which is mathematically formulated in order to capture the conflicting objectives of the \glspl{RU} in a liberalised passenger railway market like the Spanish one. 

Efficiently solving the \gls{TSA-OPEN} problem is a pivotal concern for fostering competition in the passenger railway market. In this study, we demonstrate that any criterion ${\cal A}$ employed by the \gls{IM} must select points from the Pareto front. We propose two solution approaches with these characteristics: the priority rule and the equity rule. Both criteria are straightforward to implement using linear programming and entail low computational costs. Additionally, we have introduced heuristic versions of these criteria in this study, driven by the necessity to compute the equilibrium state to study these rules, although these heuristics are not required for practical implementation.

Additionally, this paper presents a case study designed to evaluate the performance of the two proposed approaches. The study focuses on the Madrid-Barcelona corridor within the Spanish railway market. The objective of this experiment is to investigate the function of the \gls{IM} in time slot allocation with the aim of promoting competition between the \glspl{RU} and preventing the emergence of a monopolistic situation. Heuristic and exact algorithms have been applied to solve \gls{TSA-OPEN} problem using both priority and equity rules. 

The comparison between heuristic and exact algorithms has led to the conclusion that both approaches obtain similar solutions when \gls{TSA-OPEN} problem is solved using the priority rule. However, this approach tends to benefit the operator with the highest priority, exhibiting relevant differences between the revenues of the different \glspl{RU}. 

Furthermore, we have found that the heuristic equity rule achieves a fair equilibrium, wherein indistinguishable players obtain identical revenue. This demonstrates that the use of this rule by the \gls{IM} assigns it a neutral role in the market. Additionally, we have observed that the heuristic rule yields results that deviate significantly from the optimal solutions obtained through the exact method.

The discussion of question {\bf R3} leads us to speculate that the exact equity rule also attains a fair equilibrium. However, the fact that the exact method produces deviations closer to the initial requests of the various companies makes it more practical. It allows the \glspl{RU} to anticipate the outcomes of their actions and reach the equilibrium state more quickly.

The dynamics in a liberalised railway passenger system are complex and require novel models to capture. For future works, it is intended to extend this model in order to represent the complete dynamics in a liberalised railway market.  It will lead to a more complex game-theoretic model which represents competition in liberalised passenger railway markets, taking into account optimal strategies, demand modelling and equilibrium prices.

%In this paper, we present a numerical comparison of two schemes proposed. The study is carried on the Spanish high-speed passenger rail market within the Madrid-Barcelona corridor.  The primary objective of this study is to provide a preliminary estimate of the significance of the criterion employed for ${\cal A}$ in the state of market equilibrium. To achieve this objective, we are constructing a highly simplified Nash Equilibrium model, in which all rail services are identical and passengers select  the service that is nearest to their intended travel destination. Subsequently, we will conduct multiple simulations on this model, to compute the equilibrium state for both schemes ${\cal A}$, and compare the outcomes that we derive.

\section*{Acknowledgements}
%\vspace{-0.5\baselineskip}
This work was supported by grant PID2020-112967GB-C32 funded by MCIN/AEI/10.13039/501100011033 and by {\sl ERDF A way of making Europe}.

\bibliographystyle{tfcad}
\bibliography{references}

\end{document}